\documentclass[a4paper,onecolumn,11pt, accepted=2022-01-13]{quantumarticle}
\pdfoutput=1
\usepackage[utf8]{inputenc}
\usepackage[english]{babel}
\usepackage[T1]{fontenc}
\usepackage{amsmath}
\usepackage{hyperref}
\usepackage{caption}
\usepackage{subcaption}
\usepackage{tikz}
\usetikzlibrary{spy}
\usepackage[utf8]{inputenc} 
\usepackage[T1]{fontenc}
\usepackage{mathtools}

\usepackage{listings}
\usepackage{url}
\usepackage{ifthen}
\usepackage{cite}
\usepackage{hyperref}
\usepackage{amssymb}
\usepackage{xifthen,xparse}
\usepackage{bm}
\usepackage{balance}
\usepackage{tikz,pgfplots}
\pgfplotsset{compat=newest}
\pgfplotsset{plot coordinates/math parser=false}
\usetikzlibrary{decorations.markings,calc,positioning,matrix}

\newif\iffull
\newcommand{\ket}[1]{\left\lvert #1 \right\rangle}
\newcommand{\bra}[1]{\left\langle #1 \right\rvert}
\NewDocumentCommand\ketbra{+m+g}{%
  \IfNoValueTF{#2}
    {\left\lvert #1 \right\rangle \left\langle #1 \right\vert}
  {\left\lvert #1 \right\rangle \left\langle #2 \right\rvert}%
}

\NewDocumentCommand\braket{+m+g}{%
  \IfNoValueTF{#2}
    {\left\langle #1 \vert #1 \right\rangle}
  {\left\langle #1 \vert #2 \right\rangle}%
}

\usepackage{amsmath}
\DeclareMathOperator*{\argmax}{argmax} % thin space, limits underneath in displays

\usepackage{overpic}

\newtheorem{thm}{Theorem}

\newcommand{\comment}[1]{}

\usepackage{xcolor}
\definecolor{purp}{RGB}{160, 32, 240}
\definecolor{lightblue}{RGB}{66, 134, 244}

\begin{document}
\title{Reinforcement Learning with Neural Networks for Quantum Multiple Hypothesis Testing} 

% %%% Single author, or several authors with same affiliation:
% \author{%
%   \IEEEauthorblockN{Stefan M.~Moser}
%   \IEEEauthorblockA{ETH Zürich\\
%                     ISI (D-ITET)\\
%                     CH-8092 Zürich, Switzerland\\
%                     Email: moser@isi.ee.ethz.ch}
% }

%%% Several authors with up to three affiliations:
%\author{%
%  \IEEEauthorblockN{Narayanan Rengaswamy, Robert Calderbank and Henry D. Pfister}
%  \IEEEauthorblockA{Department of Electrical and Computer Engineering\\
%                    Duke University\\
%                    Durham, North Carolina 27708, USA\\
%                    Email: \{narayanan.rengaswamy, robert.calderbank, henry.pfister\}@duke.edu}
%  \and
%  \IEEEauthorblockN{Swanand Kadhe}
%  \IEEEauthorblockA{Department of ECE\\
%                    Texas A\&M University\\ 
%                    College Station, Texas 77843, USA\\
%                    Email: swanand.kadhe@tamu.edu}
%}

%%% Many authors with many affiliations:

\author{Sarah Brandsen}
\affiliation{Department of Physics, 
          Duke University, 
          Durham, North Carolina 27708, USA.}
\email{sarah.brandsen@duke.edu}

\author{Kevin D. Stubbs}
\affiliation{Department of Mathematics,
          Duke University, 
          Durham, North Carolina 27708, USA}
\author{Henry D. Pfister}
\affiliation{Department of Mathematics,
          Duke University, 
          Durham, North Carolina 27708, USA}
\affiliation{Department of Electrical and Computer Engineering,
          Duke University,
          Durham, North Carolina 27708, USA}

%  \author{%
%   \IEEEauthorblockN{Sarah Brandsen$^{*}$,
  
%                      Kevin D. Stubbs,
%                      and Henry D. Pfister}%
%   \thanks{S. Brandsen is with the 
%           Department of Physics, 
%           Duke University, 
%           Durham, North Carolina 27708, USA.
%         H. D. Pfister is with the
%           Department of Electrical and Computer Engineering,
%           Duke University,
%           Durham, North Carolina 27708, USA.
%           H.D. Pfister and K. D. Stubbs are with the
%           Department of Mathematics,
%           Duke University, 
%           Durham, North Carolina 27708, USA.
%           Correspondence Email: sarah.brandsen@duke.edu%
%   }}

\maketitle

%%%%%
% Abstract: 
% If your paper is eligible for the student paper award, please add
% the comment "THIS PAPER IS ELIGIBLE FOR THE STUDENT PAPER
% AWARD." as a first line in the abstract. 
% For the final version of the accepted paper, please do not forget
% to remove this comment!
%
\begin{abstract}
%  THIS PAPER IS ELIGIBLE FOR THE STUDENT PAPER AWARD.  Quantum error-correcting codes can be used to protect qubits involved in quantum computation. 
Reinforcement learning with neural networks (RLNN) has recently demonstrated great promise for many problems, including some problems in quantum information theory. In this work, we apply RLNN to quantum hypothesis testing and determine the optimal measurement strategy for distinguishing between multiple quantum states $\{ \rho_{j} \}$ while minimizing the error probability. In the case where the candidate states correspond to a quantum system with many qubit subsystems, implementing the optimal measurement on the entire system is experimentally infeasible. 

We use RLNN to find locally-adaptive measurement strategies that are experimentally feasible, where only one quantum subsystem is measured in each round. We provide numerical results which demonstrate that RLNN successfully finds the optimal local approach, even for candidate states up to 20 subsystems. We additionally demonstrate that the RLNN strategy meets or exceeds the success probability for a modified locally greedy approach in each random trial.

While the use of RLNN is highly successful for designing adaptive local measurement strategies, in general a significant gap can exist between the success probability of the optimal locally-adaptive measurement strategy and the optimal collective measurement. We build on previous work to provide a set of necessary and sufficient conditions for collective protocols to strictly outperform locally adaptive protocols. We also provide a new example which, to our knowledge, is the simplest known state set exhibiting a significant gap between local and collective protocols. This result raises interesting new questions about the gap between theoretically optimal measurement strategies and practically implementable measurement strategies.
\end{abstract}

\section{Introduction}
\label{sec:intro}
Optimal quantum hypothesis testing consists of finding the quantum measurement $\{\Pi_{j}\}|_{j=1}^{m}$ to optimally distinguish between $m$ candidate states $\{\rho_{j}\}_{j=1}^{m}$ with prior probabilities $\{q_{j}\}_{j=1}^{m}$. For example, this can be used to discriminate between coherent quantum states~\cite{Ferdinand_coherent} and also to decode one of $m$ codewords that has been sent through a known noisy quantum channel~\cite{Krovi_2015,Narayanan_QBP}. One important example of locally adaptive multiple hypothesis testing protocols is the Dolinar receiver, which uses an adaptive measurement scheme to distinguish between $m$ different optical signals~\cite{Dolinar_PhysRevA}.  

Although the optimal (Helstrom) measurement has a compact expression when $m=2$, the solution is more complicated for general non-binary state discrimination. In general, the optimal measurement can be written as the solution of a semidefinite programming problem~\cite{holevo1973bounds,YKL_paper}. Techniques for solving semidefinite programming then can be used to find the minimal-error measurement and compute the optimal success probability~\cite{Kiilerich_2018, Konig_2009}. 

When the candidate states are high-dimensional (corresponding to a quantum system composed of many qubit subsystems), it can be experimentally difficult to implement operations on all subsystems at once. Thus, we also focus on finding optimal (or near-optimal) approaches that include the experimentally necessary property of locality, where only a single subsystem is measured in each round. We know that dynamic programming can be used to find an optimal local approach~\cite{bellman1954}. However, even in the simplest case where $m=2$, the complexity grows like $O(2^{n}nQ)$, where $n$ is the number of qubit subsystems and $Q$ is the number of different local measurements considered~\cite{brandsen2019adaptive}. 

A powerful alternative tool for developing optimal adaptive protocols is reinforcement learning with neural networks (RLNN), where an agent learns an optimized protocol through repeated interaction with an environment.  While RLNN was introduced more than 20 years ago~\cite{Tesauro_20yrNN, Gordon_20yrsNN}, interest in these methods was recently rekindled by its remarkable success for Atari games~\cite{atari1, atari2}. RLNN and other machine-learning approaches have been successfully applied to a variety of problems in quantum information theory: generating error-correcting sequences~\cite{Fosel_PhysRevX, Miguel}, preparation of special quantum states~\cite{floquet1, ML_exp, stateprep_2}, setting up experimental Bell tests~\cite{ML_exp_Bell}, quantum communication~\cite{quantumcommunication}, fault-tolerant quantum computation~\cite{qcomp}, quantum control~\cite{qcontrol_1, qcontrol_2, qcontrol_3, qcontrol_4}, and nonequilibrium quantum thermodynamics~\cite{nonequilibrium}. Additionally, RLNN has been applied in the closely related topic of adaptive quantum metrology~\cite{sanders1, sanders2, sanders3, sanders4}. Motivated by these successes, in this work we use RLNN to find optimal locally-adaptive measurement protocols.

To demonstrate the effectiveness of reinforcement learning, we compare the RLNN performance to other locally adaptive and collective protocols. 
% As a baseline method, we consider a locally adaptive strategy based on min-entropy minimization. It is known that the optimal collective probability of success for distinguishing between $m$ candidate states directly related to the min-entropy of the corresponding classical-quantum state $\rho_{XB} \triangleq \sum_{j=1}^{m} p_{j} \ket{j}\bra{j} \otimes \rho_{j, B}$ \cite{Konig_2009}. Our baseline locally adaptive strategy is then to select the local measurement which minimizes the expected min-entropy of the remaining subsystems. 
In all our numerical results, the neural network meets or exceeds the probability of success achieved by a modified locally greedy approach. Additionally, for every simulation with randomly generated candidate state sets, the RLNN scheme approximately meets an upper bound corresponding to the optimal collective success probability (i.e., the optimal measurement scheme when measurements are not restricted to be local). This upper bound is found via semidefinite programming (SDP) techniques outlined in~\cite{Eldar_Semidefinite2}. The RLNN performance as a function of subsystem number is investigated, and we demonstrate that the neural network attains good performance for up to 20 subsystems. We additionally show that, for any locally adaptive method, the success probability is stable under small perturbations of the candidate state sets, such as rotation errors. 

While our numerical tests show that RLNNs are a powerful tool for calculating an optimal or near-optimal locally-adaptive strategy, we additionally provide analytical results for some specific systems.
%that characterize the performance of locally adaptive protocols.
These specific results help complete the picture regarding the optimality of locally adaptive protocols in four key regimes: pure binary state discrimination, mixed binary state discrimination, pure non-binary state discrimination, and mixed non-binary state discrimination. Prior to this work it was known that locally adaptive, projective measurement strategies are optimal for pure binary state discrimination\cite{Acin-physreva05, brandsen2019adaptive} and are in general are \emph{not} optimal for both pure and mixed nonbinary state discrimination~\cite{nonlocality_1, nonlocality_2}. 
In contrast to the pure state case, few analytical results are known for mixed binary state discrimination. Previous work had shown that certain fixed strategies were not optimal for mixed binary states~\cite{Croke3} and had suggested via numerics~\cite{Higgins} that any quantized locally adaptive strategy may be suboptimal. To our knowledge, we are the first to prove this analytically. Moreover, while previous work discusses a potential gap in the case of multiple subsystems where both states are noisy, we demonstrate that a gap can exist for \emph{any} nontrivial subsystem number and only one mixed state.

Still, the main significance of our paper lies in applying novel machine learning methods to quantum state discrimination. To our knowledge, we provide the first algorithm capable of successfully distinguishing between an arbitrary candidate state set via locally adaptive protocols. Our result demonstrates that not only is the optimal locally adaptive measurement strategy unknown for general state discrimination problems, but additionally the optimal locally adaptive success probability is also unknown. This further motivates the need for a reliable algorithm which can always find the optimal or close-to-optimal locally adaptive strategy, which our RLNN provides. Such an algorithm would be of broad interest not only to researchers working in the field of quantum hypothesis testing, but also to any researchers who are applying machine learning techniques to similar problems in other areas of physics.

\section{Applying Reinforcement Learning to Quantum Hypothesis Testing}
\label{sec:reinforcement_learning}
Each round of the reinforcement learning process involves an agent choosing one action from an allowed action space, implementing the action, and receiving a reward from the environment. For a Markov decision process, the agent can eventually learn to choose actions according to an optimal policy that maximizes the expected future reward. For the problem at hand, the agent is trained to learn the optimal adaptive measurement strategy as well as the optimal adaptive order in which subsystems should be measured.

In the context of state discrimination, the environment is a parameterized measurement protocol for the quantum system of interest. The action space (denoted by $\mathcal{A}$) is the set of allowed quantum measurements. Denote by $s_{t}$ the state of the environment just before round $t$ and let $n$ be the total number of rounds. The agent's policy, $\pi_{\theta}(a_{t}|s_{t})$, is parameterized by $\theta$ and equals the probability of selecting action $a_{t}\in\mathcal{A}$ in round $t$ conditioned on the state $s_{t}$ of the environment. The goal of training is for the agent to learn the optimal policy $\pi_{\theta}^{*}$ which maximizes a given reward function. \\

We consider the task of deriving the minimum-error adaptive measurement protocol to distinguish between $m$ tensor-product quantum states $\{\rho_{j}\}|_{j=1}^{m}$ with prior probability vector $\bm{q}$ where $q_{j} = \text{Pr}(\rho = \rho_{j})$. To reduce the number of measurement parameters and thus the size of the action space, we restrict to the case where each candidate state is real-valued. Since each candidate state is assumed to be a tensor product of $n$ subsystems, it can be written as
\begin{align*}
  \rho_{j} = \bigotimes_{k=1}^{n} \rho_{j}^{(k)},
\end{align*}
where $\rho_{j}^{(k)}$ is a qubit density matrix for all $j\in \{1, ..., m\}$ and all $k \in \{1, ..., n\}$. Thus, the quantum system $\rho$ is composed of $n$ unentangled qubit subsystems.

We build an OpenAI gym environment~\cite{OpenAi} capable of simulating local measurement protocols. In each round, the algorithm chooses the next subsystem $j$ to measure as well as which measurement to implement.  \\~\\
The action space $\mathcal{A}$ consists of elements $(\ell, k)$ where $\ell \in \{1, ..., 20\}$ selects which measurement in the allowed measurement set is to be implemented and $k \in \{1, ..., n\}$ is the subsystem to be measured. More specifically, $\ell$ corresponds to implementing the binary real qubit POVM,
\begin{align*}
    \hat{\Pi}_{Q}(\ell) & \triangleq \Bigg\{
    \begin{pmatrix} 
    \sin^{2}(\frac{\ell \pi}{2Q}) & \frac{1}{2} \sin(\frac{\ell \pi}{Q}) \\[1ex]
    \frac{1}{2} \sin(\frac{\ell \pi}{Q}) & \cos^{2}(\frac{\ell \pi}{2Q})
    \end{pmatrix}, \ 
    \begin{pmatrix} 
    \cos^{2}(\frac{\ell \pi}{2Q}) & -\frac{1}{2} \sin(\frac{\ell \pi}{Q}) \\[1ex]
    -\frac{1}{2} \sin(\frac{\ell \pi}{Q}) & \sin^{2}(\frac{\ell \pi}{2Q})
    \end{pmatrix},
    \Bigg\}
\end{align*}
and $Q=20$ (unless otherwise specified). The set of allowed measurement is $\{\hat{\Pi}_{Q}(\ell)\}|_{\ell=1}^{Q}$ which corresponds to binary real qubit POVMs spaced evenly on the Bloch sphere. For a given $Q$, this action set minimizes the worst case quantization error. Increasing the quantization beyond $Q=20$ (or allowing continuous choice of measurement) slowed the training time and did not offer observable gain in performance, so $Q=20$ was chosen as the smallest quantization which yields near optimal results. \\
For a given set of candidate states, the state set $\mathcal{S}$ consists of elements $(\mathbf{p}, \mathbf{v})$ where $\mathbf{p}$ denotes the updated probabilities for each candidate state and $\mathbf{v}$ is a length-$n$ vector which specifies which subsystems have been measured. More specifically, given starting prior $\mathbf{q}$ and measurement results $\mathbf{d}$, the updated prior is denoted by $p(\mathbf{q}, \mathbf{d})$. The list of subsystems which have been measured is given by the length-$n$ vector $\bm{v}$ where $v_{k} = 1$ if subsystem $k$ has already been measured and $0$ otherwise. Thus, the overall state of the environment, given starting prior $\mathbf{q}$ and measurement history $\mathbf{d}$, may be represented as $s \triangleq (p(\bm{q}, \bm{d}), \bm{v})$. The episode is terminated when all subsystems except one have been measured, or equivalently when $\sum_{i} \bm{v}_{i} = n-1$. \\
 When only one subsystem remains unmeasured the optimal final measurement is automatically determined through semidefinite programming. The reward is given by the probability of successfully decoding the actual state ($\rho = \rho_{j^{*}}$) after the final local measurement, where a successful decoding occurs if
\begin{align*}
    j^{*} = \argmax_{j \in \{1, ..., m\}}(p_{j}(\bm{q}, \bm{d})),
\end{align*}
where $\bm{d}$ is the vector containing all previous measurement results and $\bm{p}(\bm{q}, \bm{d})$ is the updated probability given initial prior $\mathbf{q}$ and measurement results $\bm{d}$. Additionally, in each round a penalty of $-0.3$ is given if the agent attempts to re-measure an already measured subsystem, as for qubit subsystems re-measuring an already measured subsystem is non-informative.

\section{Details of Implementation}

\begin{figure}[h!]
\begin{minipage}{.45\textwidth}
%\begin{figure}
  \begin{center}
  % This file was created by tikzplotlib v0.9.1.
\begin{tikzpicture}[scale = 0.73]

\begin{axis}[
legend cell align={left},
legend style={fill opacity=0.8, draw opacity=1, text opacity=1, at={(0.97,0.03)}, anchor=south east, draw=white!80!black},
tick align=outside,
tick pos=left,
x grid style={white!69.0196078431373!black},
xlabel={trial},
xmin=-0.45, xmax=9.45,
xtick style={color=black},
y grid style={white!69.0196078431373!black},
ylabel={$\text{P}_{\text{succ}}$},
ymin=0.58032784158982, ymax=0.971662402357947,
ytick style={color=black}
]
\addplot [only marks, green!50.1960784313725!black, mark=square*, mark size=3, mark options={solid}]
table {%
0 0.59824491153189
1 0.626085328291364
2 0.763419318288149
3 0.806701570355381
4 0.815537407124017
5 0.847699048782477
6 0.892634224708255
7 0.907649948991962
8 0.940513919284683
9 0.953874467777577
};
\addlegendentry{SDP}
\addplot [only marks, black, mark=triangle*, mark size=3, mark options={solid}]
table {%
0 0.59811577617019
1 0.625967331956533
2 0.756917924931999
3 0.806512865605547
4 0.815348703653875
5 0.847304636493439
6 0.892496169332839
7 0.906633142439835
8 0.9399553610788
9 0.953835287969143
};
\addlegendentry{RLNN}
\end{axis}
\end{tikzpicture}
  %\scalebox{0.73}{\input{helstrom_vs_NN}}
    \caption{\label{fig:tuned} Performance of tuned network versus the SDP upper bound on the optimal success probability. }
%\end{figure}
\end{center}
\end{minipage}
\begin{minipage}{.05\textwidth}

\hspace*{0.1cm}

\end{minipage}
\begin{minipage}{.45\textwidth}
%\begin{figure}
 \begin{center}
%   \scalebox{0.73}{
%   \input{fig10.tex}
%   }
% This file was created by tikzplotlib v0.9.1.
\begin{tikzpicture}[scale = 0.73]

\begin{axis}[
legend cell align={left},
legend style={fill opacity=0.8, draw opacity=1, text opacity=1, at={(0.97,0.03)}, anchor=south east, draw=white!80!black},
tick align=outside,
tick pos=left,
x grid style={white!69.0196078431373!black},
xlabel={trial},
xmin=-0.45, xmax=9.45,
xtick style={color=black},
y grid style={white!69.0196078431373!black},
ylabel={$P_{\text{succ}}$},
ymin=0.4942010205, ymax=1.0154430495,
ytick style={color=black}
]
\addplot [only marks, green!50.1960784313725!black, mark=square*, mark size=2.75, mark options={solid}]
table {%
0 0.51789384
1 0.54226194
2 0.57242864
3 0.58277788
4 0.69474426
5 0.75085468
6 0.79664274
7 0.90798945
8 0.91688166
9 0.99175023
};
\addlegendentry{Helstrom}
\addplot [only marks, black, mark=*, mark size=2, mark options={solid}]
table {%
0 0.5288
1 0.54182
2 0.57844
3 0.57536
4 0.69218
5 0.7374
6 0.78656
7 0.8856
8 0.8974
9 0.9696
};
\addlegendentry{RLNN}
\end{axis}

\end{tikzpicture}

  \caption{\label{fig:untuned} Performance of network before tuning versus the SDP upper bound on the optimal success probability. }
%\end{figure}
\end{center}
\end{minipage}
\end{figure}
 
We train the agent using the proximal policy optimization (PPO) algorithm~\cite{schulman2017proximal}. From numerical simulations, we found that a PPO algorithm significantly outperformed a DQN-based algorithm which had poor performance and training instability. Given that the environment in our problem is not too expensive to sample from, PPO worked well and we did not try DDPG or SAC, which can be more sample efficient. PPO algorithms generally train well on problems with discrete action spaces and environments that are cheap to sample from. PPO algorithms additionally offer the benefit of relatively straightforward hyperparameter tuning and hyperparameters did not need to be re-tuned for each combination of $(m,n)$.

Results are then generated using the default PPO algorithm from the RLlib package included in Ray version 0.7.3~\cite{liaw2018tune, liang2017rllib}. After hyperparameter tuning of the learning rate, we set the learning rate to be $\eta = 5 \times 10^{-5}$. For the remaining hyperparameters, we find the default parameter settings to be optimal, including the clipping parameter $\epsilon = 0.3$ and the discount factor $\gamma = 0.99$. Comparison of the tuned versus untuned training is depicted in Figures 1 and 2.

\begin{figure}[h!]
\centering
\small
\begin{overpic}[width=100mm]{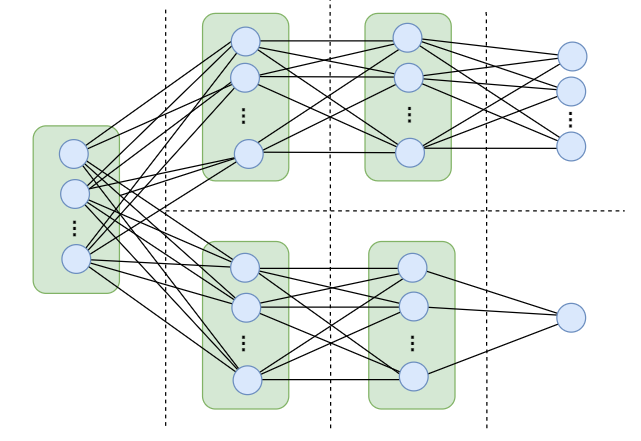}
\footnotesize
\put(-5, 37){Input}
\put(-5, 33){Layer}
\put(40, 0){Two Hidden Layers}
\put(93, 59){$\pi(a_{1}|s)$}
\put(93, 54){$\pi(a_{2}|s)$}
\put(93, 45){$\pi(a_{|\text{A}|}|s)$}
\put(93, 18){$V(s)$}
\end{overpic}
 \caption{\label{fig:nn_config} Neural Network configuration, consisting of one input layer, two parallel subnetworks. One subnetwork outputs an estimate of the value $V(s)$ for state $s$, one outputs an estimate of the policy $\pi(a|s)$.}
\centering
\end{figure}
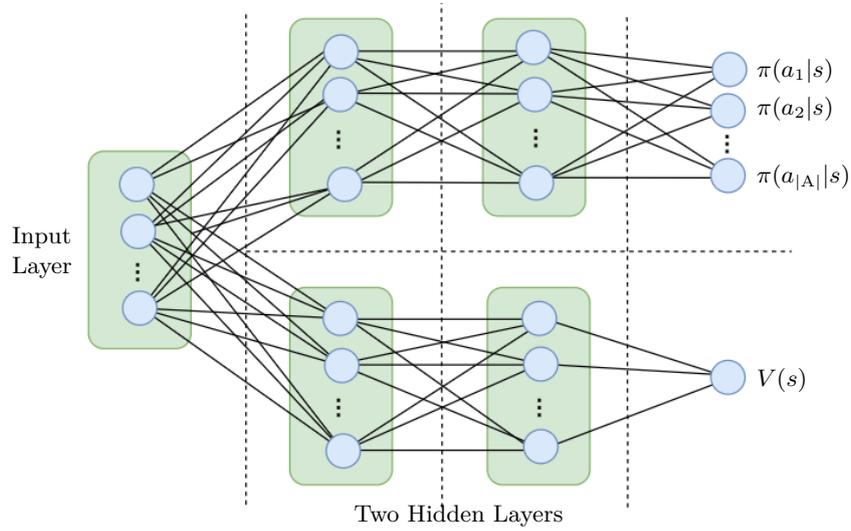

We use a fully connected neural network where the input layer (with $m+n-1$ neurons) takes the state $s$ as the input, as depicted in Fig. ~\ref{fig:LG}. This feeds into two parallel sets of subnetworks, each of which has two hidden layers of 256 neurons, tanh activation functions, and their own linear output layers. Although additional hidden layers could be added, this would increase the training time required. The output layer of the first subnetwork consists of a single neuron, and computes an estimate for the value of states. The output layer of the second subnetwork has $n Q$ neurons (i.e. the number of allowed actions) and computes the policy, $\pi(a|s)$. See
https://github.com/SarahBrandsen/RLNN-QSD for numerical results and the source code used to obtain them.

\section{Numerical Results for RLNN Performance}

As an initial benchmark of the RLNN performance, we compare it to known optimal results in several special cases. 

In the case of binary discrimination (i.e. $m=2$) between tensor products of pure states such that $\rho_{j}^{(k)} = \ket{\smash{\psi_{j}^{(k)}}}\!\!\bra{\smash{\psi_{j}^{(k)}}}$ for all $k \in \{1, ..., n\}$ and $j \in \{1, 2\}$, it has been shown that the optimal collective success probability, $\text{P}_{\text{SDP}}$ can be achieved through locally-adaptive strategies~\cite{Acin-physreva05, brandsen2019adaptive}. The collective success probability, $\text{P}_{\text{SDP}}$, is found using semidefinite programming techniques introduced by~\cite{Eldar_Semidefinite2}.
We randomly generate ten trials with $n=3$ and order the trials according to increasing distinguishability measured by $\text{P}_{\text{SDP}}$. For each trial, we compare this success probability with the RLNN success probability, $\text{P}_{\text{RLNN}}$, as shown in Fig.~\ref{fig:hels_vs_NN}. The neural network attains the correct (optimal) success probability in each case, with a very small gap that is likely due to action space quantization.

\begin{figure}[h!]

\begin{minipage}{.45\textwidth}
  \begin{center}
\begin{tikzpicture}[scale = 0.83]
\begin{axis}[
legend cell align={left},
legend style={fill opacity=0.8, draw opacity=1, text opacity=1, at={(0.97,0.03)}, anchor=south east, draw=white!80!black},
tick align=outside,
tick pos=left,
x grid style={white!69.0196078431373!black},
xlabel={trial},
xmin=-0.45, xmax=9.45,
xtick style={color=black},
y grid style={white!69.0196078431373!black},
ylabel={$\text{P}_{\text{succ}}$},
ymin=0.58032784158982, ymax=0.971662402357947,
ytick style={color=black}
]
\addplot [only marks, green!50.1960784313725!black, mark=square*, mark size=3, mark options={solid}]
table {%
0 0.59824491153189
1 0.626085328291364
2 0.763419318288149
3 0.806701570355381
4 0.815537407124017
5 0.847699048782477
6 0.892634224708255
7 0.907649948991962
8 0.940513919284683
9 0.953874467777577
};
\addlegendentry{SDP}
\addplot [only marks, black, mark=triangle*, mark size=3, mark options={solid}]
table {%
0 0.59811577617019
1 0.625967331956533
2 0.756917924931999
3 0.806512865605547
4 0.815348703653875
5 0.847304636493439
6 0.892496169332839
7 0.906633142439835
8 0.9399553610788
9 0.953835287969143
};
\addlegendentry{RLNN}
\end{axis}
\end{tikzpicture}
  \caption{\label{fig:hels_vs_NN} Probability of success for the optimal RLNN policy after 1000 training iterations vs. the optimal collective measurement for tensor-products of pure states when $m=2$, $n=3$. The neural network closely approximates the optimal success probability in each trial, with any gap likely arising from quantization of the action space. }
  \end{center}
\end{minipage}
\begin{minipage}{.05\textwidth}

\hspace*{0.1cm}

\end{minipage}
\begin{minipage}{.45\textwidth}
  \begin{center}
\begin{tikzpicture}[scale = 0.83]

\definecolor{color0}{rgb}{0.12156862745098,0.466666666666667,0.705882352941177}
\definecolor{color1}{rgb}{1,0.498039215686275,0.0549019607843137}
\definecolor{color2}{rgb}{0.172549019607843,0.627450980392157,0.172549019607843}
\definecolor{color3}{rgb}{0.83921568627451,0.152941176470588,0.156862745098039}
\definecolor{color4}{rgb}{0.580392156862745,0.403921568627451,0.741176470588235}
\definecolor{color5}{rgb}{0.549019607843137,0.337254901960784,0.294117647058824}
\definecolor{color6}{rgb}{0.890196078431372,0.466666666666667,0.76078431372549}
\definecolor{color7}{rgb}{0.737254901960784,0.741176470588235,0.133333333333333}
\definecolor{color8}{rgb}{0.0901960784313725,0.745098039215686,0.811764705882353}

\begin{axis}[
tick align=outside,
tick pos=left,
x grid style={white!69.0196078431373!black},
xlabel={training iteration},
xmin=5, xmax=200,
xtick style={color=black},
y grid style={white!69.0196078431373!black},
ylabel={$\text{P}_{\text{SDP}}- \text{P}_{\text{RLNN}}$},
ymin=-0.000624382056097389, ymax=0.0127482042493766,
ytick style={color=black}
]
\addplot [line width=0.04pt, color0, mark=*, mark size=1, mark options={solid}]
table {%
10 0.00572996423199046
20 0.0019300164832956
30 0.00104924284055785
40 0.000635746512557711
50 0.000370622163642365
60 0.000514737533201415
70 0.000261352394502978
80 0.000248425003744446
90 0.000299551482272453
100 0.000246220685388709
110 0.000254459104932381
120 0.000653389947486649
130 0.000284235299399804
140 0.000201699710748349
150 0.00261613197211141
160 0.0002810206175059
170 3.64560103077727e-05
180 4.13781539972957e-05
190 6.74381379283107e-05
200 0.000422422112299592
};
\addplot [line width=0.04pt, color1, mark=*, mark size=1, mark options={solid}]
table {%
10 0.0121403594173096
20 0.00674917156363886
30 0.00596499720426003
40 0.00479885150948645
50 0.00291632197947711
60 0.00226216040339611
70 0.000500425583243147
80 0.000682068112526246
90 0.000201579248346073
100 0.000282560529364373
110 0.000481622683291216
120 0.000350542269920373
130 0.000362222671743462
140 0.000240904761000471
150 0.000402465758069237
160 0.000126005719741085
170 0.000388793662289943
180 0.00302947970896039
190 0.000111162751196936
200 0.000130771465869484
};
\addplot [line width=0.04pt, color2, mark=*, mark size=1, mark options={solid}]
table {%
10 0.0028703430497502
20 0.000860097389411507
30 0.000472772945097621
40 0.000316785454525403
50 0.000118480120284592
60 0.000301043471581774
70 8.7230627581758e-05
80 9.75541651332401e-05
90 7.72442645016191e-05
100 8.91466959982434e-05
110 0.000196714985017277
120 0.000142918607866638
130 0.000318538929216072
140 5.04533267348117e-05
150 4.35731201616774e-05
160 9.47963175099709e-05
170 0.000113064672610075
180 0.000121109381294393
190 5.54836302626427e-05
200 5.02163637038366e-05
};
\addplot [line width=0.04pt, color3, mark=*, mark size=1, mark options={solid}]
table {%
10 0.00541971812692943
20 0.00179405628842588
30 0.00129451508146095
40 0.00115130609622016
50 0.00114506691930294
60 0.00130624774769428
70 0.000610573026494698
80 0.000682084183506682
90 0.000291199615136462
100 0.000565252897724466
110 0.000280699016182107
120 0.000927332278061521
130 0.000123712694914957
140 0.000189665482439927
150 9.73266994588329e-05
160 0.000196203845778342
170 0.000287771905525203
180 0.0016628865280548
190 0.000155046323897734
200 0.000100639252331058
};
\addplot [line width=0.04pt, color4, mark=*, mark size=1, mark options={solid}]
table {%
10 0.0112214142410797
20 0.0058726146188951
30 0.00421665982735653
40 0.00374049214461603
50 0.00326018327044797
60 0.00222557852620842
70 0.00252739486422948
80 0.00118202795043099
90 0.00137396259178191
100 0.000878658938802013
110 0.000665236940150971
120 0.000668434666222906
130 0.000415866497659678
140 0.000498631243908254
150 0.000619718126400159
160 0.000322597925715762
170 0.000564028981781539
180 0.000458276543114366
190 0.000439857530547982
200 0.000394123179665695
};
\addplot [line width=0.04pt, color5, mark=*, mark size=1, mark options={solid}]
table {%
10 0.00499095627468793
20 0.00164337505995282
30 0.00110504759249219
40 0.000709760848361385
50 0.000405279507551648
60 0.000550514077270581
70 0.000456067784130987
80 0.000347792697728444
90 0.000823523826529282
100 0.000827137730349481
110 0.000593599223883468
120 0.000530044810801988
130 4.25739220630827e-05
140 0.0010341337798222
150 0.00057260134686421
160 0.000453932315269134
170 0.000461343030472294
180 0.000750664719824545
190 0.000921366124094436
200 0.000279733779643299
};
\addplot [line width=0.04pt, color6, mark=*, mark size=1, mark options={solid}]
table {%
10 0.00435678564656028
20 0.00118974751322087
30 0.000659701539742374
40 0.000673809293284311
50 0.000340733556222661
60 0.000285611419850063
70 0.000207320064845362
80 0.000263341474224021
90 0.000162217915208096
100 0.000151126848375038
110 0.000161184985314811
120 3.93198094058578e-05
130 0.000113189865863728
140 9.13613833327664e-05
150 0.00024190258716672
160 0.00014064099646216
170 0.000333330976245882
180 2.87851720165433e-05
190 0.000332651053110977
200 0.000280521872632278
};
\addplot [line width=0.04pt, white!49.8039215686275!black, mark=*, mark size=1, mark options={solid}]
table {%
10 0.00839139709082004
20 0.00256321781155655
30 0.00232318139902343
40 0.00146986459454146
50 0.00136229966256685
60 0.00214011576339856
70 0.00118183909744252
80 0.00128689699197038
90 0.000836119899017218
100 0.000666722789309127
110 0.000538929756653261
120 0.00171630951148516
130 0.000693705900255082
140 0.000767542200229676
150 0.000655463188587779
160 0.000135053351809411
170 0.00049764685708531
180 0.000334771631766206
190 0.000366701954876203
200 0.000466459997209068
};
\addplot [line width=0.04pt, color7, mark=*, mark size=1, mark options={solid}]
table {%
10 0.00995296684030589
20 0.00548442507231239
30 0.00393319535168368
40 0.00418220530342273
50 0.00408127414233306
60 0.00245320779382918
70 0.00227078055333929
80 0.002217617979132
90 0.00206060788988249
100 0.00110284114100789
110 0.00108526923571839
120 0.00113155887254668
130 0.00101864383167805
140 0.00133396373127759
150 0.00118628329824544
160 0.00144551279996208
170 0.00117016087799338
180 0.0019227585479773
190 0.00105655513252434
200 0.00154885317898978
};
\addplot [line width=0.04pt, color8, mark=*, mark size=1, mark options={solid}]
table {%
10 0.0103418779432091
20 0.00628989170664096
30 0.00601555091909756
40 0.00532693985242638
50 0.0035280169463956
60 0.00247078670314138
70 0.00173868123522114
80 0.00205380777958664
90 0.000600678710115266
100 0.000702513278205918
110 0.000896739463048801
120 0.00141177960835959
130 0.00156416786152269
140 0.000932443634741209
150 0.00190239619054455
160 0.00121837101205846
170 0.00139148654062171
180 0.00122379161403707
190 0.0012114192094026
200 0.000872711082198441
};
\end{axis}
 \end{tikzpicture}
  \caption{\label{fig:n5_training} Difference between RLNN reward and Helstrom (SDP) success probability as a function of training iteration. We observe that the RLNN success probability stabilizes after 100 training iterations, with occasional fluctuations.}
%\end{figure}
\end{center}
\end{minipage}
\end{figure}

\begin{figure}
 \begin{center}
 \begin{tikzpicture}[scale = 0.83]
\begin{axis}[
legend cell align={right},
legend entries={{Optimal},{RLNN}},
legend style={at={(0.37,0.03)}, anchor=south east, draw=white!80.0!black},
tick align=outside,
tick pos=left,
x grid style={white!69.01960784313725!black},
xlabel={$n$},
xmajorgrids,
xmin=1, xmax=8,
y grid style={white!69.01960784313725!black},
ylabel={$\text{P}_{\text{succ}}$},
ymajorgrids,
ymin=0.25, ymax=0.9
]
% \addlegendimage{no markers, blue}
% \addlegendimage{no markers, black}
\addplot [semithick, green!50.1960784313725!black, mark=square*, mark size=2, mark options={solid}]
table [row sep=\\]{%
1   0.85 \\
2   0.725 \\
3   0.614125 \\
4   0.52200625 \\
5   0.443705 \\
6   0.37714951562 \\
7   0.3205770882812499\\
8   0.27249052 \\
};
\addplot [semithick, black, mark=*, mark size=1.5, mark options={solid}]
table [row sep=\\]{%
1	0.85 \\
2   0.7225 \\
3   0.614125 \\ 
4   0.521685 \\
5   0.443549788 \\
6   0.377135 \\
7   0.32048 \\
8   0.271857 \\
};
\end{axis}

\end{tikzpicture}
  \caption{\label{fig:Sasaki} Performance of the RLNN policy after 150 training iterations vs. the optimal success probability as a function of the number of subsystems $n$. The RLNN approach converges to the optimal local approach in this example for $1 \leq n \leq 8$.} \vspace{-2mm}
  \end{center}
\end{figure}
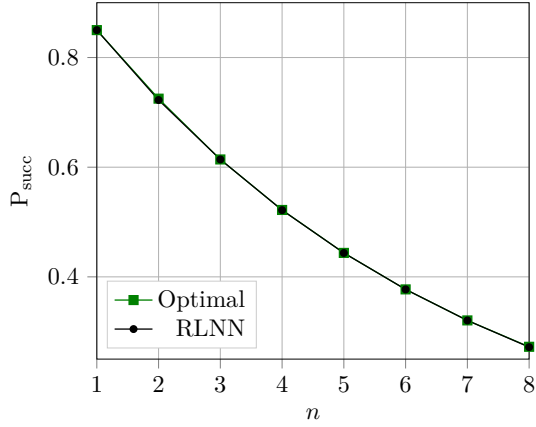

An additional case where locally adaptive protocols are strictly optimal has been found by Sasaki et. al in~\cite{Sasaki_2008}. Consider a set of states $\mathcal{S}_{1} \triangleq \{\rho_{j}\}|_{j=1}^{m}$ and associated probabilities $\{q_{j}\}|_{j=1}^{m}$. Suppose the known optimal POVM for these is $\{\Pi_{j}\}|_{j=1}^{m}$. The set of $n$-subsystem product states generated by $\mathcal{S}$ can be written as
\begin{align*}
    \mathcal{S}_{n} \triangleq \bigg\{ \bigotimes_{j=1}^{n} \rho_{i_{j}} \ \Big| \ i \in \{1, ..., m\}^{n} \bigg\},
\end{align*}
with corresponding probabilities defined as $q_{i_{1}...i_{n}} \triangleq q_{i_{1}} \times ... \times q_{i_{n}}$. Then, the optimal POVM candidate state set $\mathcal{S}_{n}$ has elements that can be written in tensor product form as:
\begin{align*}
    \Pi_{i_{1}...i_{n}} = \bigotimes_{j=1}^{n} \Pi_{i_{j}}. 
\end{align*}
 This provides a useful test of the neural network performance. We take the initial state set to be $\mathcal{S}_{1} = \{\rho_{1}, \rho_{2}\}$, where
 $\rho_{1} = \begin{psmallmatrix} 0.85 & 0 \\ 0 & 0.15 \end{psmallmatrix}$ and $\rho_{2} = \begin{psmallmatrix} 0.15 & 0 \\ 0 & 0.85 \end{psmallmatrix}$. Since the optimal local POVM belongs to the allowed action set, there should be no quantization loss. We train the neural network for 1000 iterations (using a custom learning rate schedule where the learning rate starts at $5.5 \times 10^{-5}$ and decays by $0.95$ every 10 iterations), and compare the neural network performance after training to the optimal success probability. For $1 \leq n \leq 8$, the neural network attains or approximately attains the exact success probability, as depicted in Fig.~\ref{fig:Sasaki}.

\section{Comparison to SDP-based Locally Adaptive Strategies}

Just as the collective SDP measurement provides an upper bound for the optimal locally adaptive success probability, simple locally adaptive algorithms such as locally greedy algorithms provide a lower bound. In this section, we introduce a local SDP-based approach, and demonstrate numerically that the RLNN always meets or exceeds the success probability of the local SDP-based approach.

In the case of binary state discrimination, locally greedy algorithms are optimal for pure states and close-to-optimal for mixed states. Our choice of the local SDP-based algorithm as a ``good'' simple strategy is motivated by the fact that it reduces to a locally greedy protocol when $m=2$. Additionally, for $m>2$, we found through numerical simulation that the local SDP-based based approach generally performs better than locally greedy protocols. Finally, we compare the RLNN algorithm to the local SDP-based algorithm via simulations with $n=3$ and $n=4$ and demonstrate that the RLNN always meets or significantly exceeds the local SDP-based success probability.

The SDP-based local algorithm selects the local measurement which maximizes the expected (collective) success probability of future rounds.  Let $\mathcal{S}$ be the set of remaining unmeasured subsystems. For each round, the algorithm chooses to measure the subsystem $l \in \mathcal{S}$ and implement the measurement $a$ such that 
\begin{align*}
    (a, l) &= \argmax_{(a, \ell) \in \mathcal{A} \times \mathcal{S}}  \sum_{d'=0}^{|a|} \ \text{Pr}(d_{n-|\mathcal{S}\backslash \ell|} = d') \\
    & \hspace{7em} \times \text{P}_{\text{succ, coll}}\bigg(\{ \rho_{j}^{\mathcal{S} \backslash l} \} \ \big | \ q, d_{n-|\mathcal{S}\backslash \ell|} = d' \bigg), 
\end{align*}
where $\text{P}_{\text{succ, coll}}\big(\{ \rho_{j}^{\mathcal{S} \backslash l} \} \ \big | \ q, d_{n-|\mathcal{S}\backslash \ell|} = d' \big)$ equals the success probability of implementing an optimal collective measurement on the remaining subsystems (with indices belonging to set $\mathcal{S} \backslash l$), given the prior for round $j$ was $q$ and the outcome of action $a$ was $d'$. 

% A necessary and sufficient condition for optimality for a given set $\{\rho_{j}\}$  is that for all possible priors $q$ and all $ k \in [1, ..., n]$:
% \begin{align*}
%      \argmax_{a \in \text{A}}  \mathbb{E} \bigg[ \text{P}_{\text{succ, coll}}\bigg(\{ \rho_{j}^{(k+1:n)} \} \ \big | \ q, d_{k} = \ell \bigg) \bigg] =  \argmax_{a \in \text{A}}  \mathbb{E} \bigg[ \text{P}_{\text{succ, locally \ adpative}}\bigg(\{ \rho_{j}^{(k+1:n)} \} \ \big | \ q, d_{k} = \ell \bigg) \bigg],
%  \end{align*}
%  namely, that the action which optimizes future collective success probability also optimizes future locally adaptive success probability.  \\
In the special case where $m=n=3$, all candidate states are pure states, and all subsystems identical copies, the performance of the SDP-based local algorithm and the RLNN algorithm appear to be identical for each of 5 random trials, as depicted in Fig. ~\ref{fig:SDP_local_RLNN}.  However, we demonstrate that simpler locally adaptive strategies such as the min-entropy approach are not sufficient to find the optimal locally adaptive strategy, as when $n=4$ and the candidate states are mixed, a significant gap appears between the RLNN results and the SDP-based local algorithm, as shown in Fig. ~\ref{fig:SDP_local_RLNN5}.

\begin{figure}[h!]
\begin{minipage}{.45\textwidth}
%\begin{figure}
  \begin{center}
\begin{tikzpicture}[scale = 0.83]

\begin{axis}[
legend cell align={left},
legend style={fill opacity=0.8, draw opacity=1, text opacity=1, at={(0.97,0.03)}, anchor=south east, draw=white!80!black},
tick align=outside,
tick pos=left,
x grid style={white!69.0196078431373!black},
xlabel={trial},
xmin=-0.2, xmax=4.2,
xtick style={color=black},
y grid style={white!69.0196078431373!black},
ylabel={$\text{P}_{\text{succ}}$},
ymin=0.498797971393737, ymax=0.964951676694631,
ytick style={color=black}
]
\addplot [only marks, green!50.1960784313725!black, mark=square*, mark size=2, mark options={solid}]
table {%
0 0.535937713621666
1 0.713190729051263
2 0.754666745697288
3 0.771182297670398
4 0.943762871908227
};
\addlegendentry{SDP}
\addplot [only marks, black, mark=*, mark size=2, mark options={solid}]
table {%
0 0.521937068987506
1 0.700296529676741
2 0.752244224543899
3 0.757314563659303
4 0.930963368430915
};
\addlegendentry{RLNN}
\addplot [ only marks, red, mark=x, mark size=2, mark options={solid}]
table {%
0 0.519986776180141
1 0.700262401802845
2 0.75199047851289
3 0.757619056717751
4 0.929029335507669
};
\addlegendentry{local SDP-based}
\end{axis}

\end{tikzpicture}

  \caption{\label{fig:SDP_local_RLNN} Plot of success probability for RLNN after 250 training iterations, the collective optimal (SDP) measurement, and the SDP-based local algorithm, over 5 trials with $m=3$, $n=3$ and pure states.}
\end{center}
\end{minipage}
\begin{minipage}{.05\textwidth}

\hspace*{0.1cm}

\end{minipage}
\begin{minipage}{.45\textwidth}
\begin{center}
 \begin{tikzpicture}[scale = 0.83]

\begin{axis}[
legend cell align={left},
legend style={fill opacity=0.8, draw opacity=1, text opacity=1, at={(0.55,0.6)}, anchor=south east, draw=white!80!black},
tick align=outside,
tick pos=left,
x grid style={white!69.0196078431373!black},
xlabel={trial},
xmin=0.8, xmax=5.3,
xtick style={color=black},
y grid style={white!69.0196078431373!black},
ylabel={$\text{P}_{\text{succ}}$},
ymin=0.5, ymax=0.79,
ytick style={color=black}
]
\addplot [only marks, green!50.1960784313725!black, mark=square*, mark size=2, mark options={solid}]
table {%
1 0.569489173839213
2 0.609676180868385
3 0.622075689217042
4 0.622131903992726
5 0.75525123656342
};
\addlegendentry{SDP}
\addplot [only marks, black, mark=*, mark size=2, mark options={solid}]
table {%
1 0.567243172241975
2 0.603210254289557
3 0.617674797791913
4 0.60954620867448
5 0.736016198082895
};
\addlegendentry{RLNN}
\addplot [only marks, red, mark=x, mark size=3, mark options={solid}]
table {%
1 0.55105291829008
2 0.589311370368079
3 0.621505043345545
4 0.602821217316382
5 0.736307796964912
};
\addlegendentry{local SDP-based}
\end{axis}

\end{tikzpicture}
  \caption{\label{fig:SDP_local_RLNN5} Plot of success probability for RLNN after 250 training iterations, the collective optimal (SDP) measurement, and the SDP-based local algorithm, over 5 trials with $m=3$, $n=5$. }
  \end{center}
\end{minipage}
\end{figure}

\section{ Pure State Discrimination}

In the special case of binary state discrimination ($m=2$), it has been shown~\cite{Acin-physreva05, brandsen2019adaptive} that locally-greedy algorithms are optimal for distinguishing between pure tensor product states. Thus, for pure binary state discrimination, the success probability of the optimal collective measurement can be achieved through a simple locally-greedy algorithm. It has additionally been demonstrated that for $m \geq 3$, there exist simple state sets such that the optimal locally adaptive algorithm performs worse than the optimal collective measurement~\cite{Croke2017}. \\~\\
We now build on these results to determine whether, for a given $m$ and $n$, there exists a candidate state set such that there exists a significant gap between the optimal locally adaptive and optimal collective measurement.  
\begin{thm}
Denote by $\text{P}_{\mathrm{loc}}(\{\rho_{j}\}, \bm{q})$ the optimal probability of success using locally adaptive projective measurements for candidate state set $\{\rho_{j}\}$ with prior probability vector $\bm{q}$. Likewise, denote by $P_{\mathrm{coll}}(\{\rho_{j}\}, \bm{q})$ the success probability for the optimal collective measurement on the full quantum system. Then for a given $m$ and a given $n>1$, there exists at least one set of tensor product states $\{\rho_{j} = \bigotimes_{k=1}^{n} \rho_{j}^{(k)} \}|_{j=1}^{m}$ and some starting prior $\bm{q}$ such that $P_{\mathrm{loc}}(\{\rho_{j}\}, \bm{q}) < P_{\mathrm{coll}}(\{\rho_{j}\}, \bm{q})$ if and only if at least one of the following conditions is met: 
\begin{enumerate}
    \item There are more than two candidate states ($m >2$)
    \item At least one candidate state is not a pure state.
\end{enumerate}
\end{thm}
{\em Proof Sketch:} The full proof is listed in Appendix A. As a proof sketch, we note that the ``only if'' direction of the proof follows immediately from~\cite{Virmani_2001}, where it was shown that locally adaptive methods are optimal for any binary pure state discrimination problem (including discrimination problems involving entangled states.) \\~\\
To show the ``if'' direction, we introduce a simple binary state discrimination with two qutrit subsystems and demonstrate that a significant gap exists for the following candidate states:
\begin{align*}
    \rho_{+} & \triangleq \Big( \frac{1}{2} \ket{0}\!\!\bra{0} + \frac{1}{2} \ket{1}\!\!\bra{1} \Big) \otimes \Big( \frac{1}{2} \ket{0}\!\!\bra{0} + \frac{1}{2} \ket{1}\!\!\bra{1} \Big) \\
    \rho_{-} & \triangleq \frac{1}{3} \Big( \sum_{j=0}^{2} \ket{j}\otimes \ket{j} \big) \big( \sum_{k=0}^{2} \bra{k}\otimes \bra{k} \Big). 
\end{align*}
The ``if'' direction is completed by previous results which demonstrate a significant gap can exist even in the case of three two-qubit candidate states~\cite{Croke2017}. 

\section{Gap between Locally Optimal Algorithm and Collective Measurement}
\label{sec:SDP}
Finally, we use RLNN to estimate the gap between the best locally adaptive algorithm and the optimal collective (non-local) measurement in more general cases where the best locally adaptive algorithm is not otherwise known. 

The simulation setup for a given $m$ and $n$ is as follows: for each trial, we randomly generate pure tensor product candidate states and then apply depolarizing noise with a randomly chosen noise parameter. The RLNN algorithm is independently trained 5 times over 2000 iterations, and the average final success probability is compared (with error bars) to the optimal collective success probability found via SDP. Results are plotted for $m=2$, $n=3$ in Figure~\ref{fig:m2n3} and for $m=3$, $n=3$ in Figure~\ref{fig:m3n3}, and indicate that the gap between local and collective measurements increases with $m$.

\begin{figure}
\begin{minipage}{.45\textwidth}
\begin{center}
\begin{tikzpicture}[scale = 0.83]

\begin{axis}[
legend cell align={left},
legend style={fill opacity=0.8, draw opacity=1, text opacity=1, at={(0.97,0.03)}, anchor=south east, draw=white!80!black},
tick align=outside,
tick pos=left,
x grid style={white!69.0196078431373!black},
xlabel={trial},
xmin=0.8, xmax=5.2,
xtick style={color=black},
y grid style={white!69.0196078431373!black},
ylabel={$\text{P}_{\text{succ}}$},
ymin=0.576042863862518, ymax=0.867736836166256,
ytick style={color=black}
]
\path [draw=black]
(axis cs:0,0.589301680785415)
--(axis cs:0,0.589301921903891);

\path [draw=black]
(axis cs:1,0.647888093698619)
--(axis cs:1,0.647889737305162);

\path [draw=black]
(axis cs:2,0.684622560515357)
--(axis cs:2,0.684622600246292);

\path [draw=black]
(axis cs:3,0.724145586920997)
--(axis cs:3,0.724146917433);

\path [draw=black]
(axis cs:4,0.854311665744311)
--(axis cs:4,0.854311730975645);

% \addplot [semithick, black, mark=-, mark size=5, mark options={solid}, only marks]
% table {%
% 1 0.589301680785415
% 2 0.647888093698619
% 3 0.684622560515357
% 4 0.724145586920997
% 5 0.854311665744311
% };

% \addplot [semithick, black, mark=-, mark size=5, mark options={solid}, only marks]
% table {%
% 1 0.589301921903891
% 2 0.647889737305162
% 3 0.684622600246292
% 4 0.724146917433
% 5 0.854311730975645
% };

\addplot [only marks, green!50.1960784313725!black, mark=square*, mark size=3, mark options={solid}]
table {%
1 0.589586180301881
2 0.648390540973291
3 0.684716826104031
4 0.724281216130612
5 0.854478019243359
};
\addlegendentry{SDP}
\addplot [only marks, black, mark=triangle*, mark size=3, mark options={solid}]
table {%
1 0.589301801344653
2 0.647888915501891
3 0.684622580380825
4 0.724146252176998
5 0.854311698359978
};
\addlegendentry{RLNN}
\end{axis}
\end{tikzpicture}

  \caption{\label{fig:m2n3} Probability of success for SDP and RLNN when $m=2$ and $n=3$. For each trial, the RLNN success probability is computed by separately training the neural network five times with 2000 iterations each. The error bars, were they visible, would represent the standard deviation in the final success probability over the five independent trainings. But in all trials, the error bars have collapsed to nothing, and the gap between local and non-local measurements is very small.}
\end{center}
\end{minipage}
\begin{minipage}{.05\textwidth}

\hspace*{0.1cm}

\end{minipage}
\begin{minipage}{.45\textwidth}
%\begin{figure}
\begin{center}
 % This file was created by tikzplotlib v0.9.1.
\begin{tikzpicture}[scale = 0.83]

\begin{axis}[
tick align=outside,
tick pos=left,
x grid style={white!69.0196078431373!black},
xlabel={trial},
xmin=0.8, xmax=5.2,
legend cell align={left},
legend style={fill opacity=0.8, draw opacity=1, text opacity=1, at={(0.97,0.03)}, anchor=south east, draw=white!80!black},
xtick style={color=black},
y grid style={white!69.0196078431373!black},
ylabel={$\text{P}_{\text{succ}}$},
ymin=0.459277016109028, ymax=0.753833602321057,
ytick style={color=black}
]
\path [draw=black]
(axis cs:0,0.472665951845938)
--(axis cs:0,0.472666767218265);

\path [draw=black]
(axis cs:1,0.516857762989347)
--(axis cs:1,0.516858183774041);

\path [draw=black]
(axis cs:2,0.539356556187692)
--(axis cs:2,0.539356699521926);

\path [draw=black]
(axis cs:3,0.662041153234538)
--(axis cs:3,0.662041621602968);

\path [draw=black]
(axis cs:4,0.730556323900145)
--(axis cs:4,0.730560087473176);

\addplot [only marks, green!50.1960784313725!black, mark=square*, mark size=3, mark options={solid}]
table {%
1 0.473130902296789
2 0.519428210950624
3 0.547382811369203
4 0.667452149906034
5 0.740444666584147
};
\addlegendentry{SDP}
\addplot [only marks, black, mark=triangle*, mark size=3, mark options={solid}]
table {%
1 0.472666359532101
2 0.516857973381694
3 0.539356627854809
4 0.662041387418753
5 0.730558205686661
};
\addlegendentry{RLNN}
\end{axis}

\end{tikzpicture}

  \caption{\label{fig:m3n3} Probability of success for SDP and RLNN after 2000 training iterations when $m=3$, $n=3$. For each trial, the RLNN success probability is computed by separately training the neural network five times with 2000 iterations each. In all trials, the error bars have collapsed to nothing and the gap between local and non-local measurements is very small. Compared to the case where $m=2$, there is a slightly larger gap between local and non-local measurements.}
\end{center}
\end{minipage}
\end{figure}

\section{Performance for a Large Number of Subsystems}
\label{sec:Large_N}
We examine how the RLNN performance varies as a function of $n$, and demonstrate good performance for up to $n=10$ subsystems. However, for $n \geq 20$, the RLNN begins to have suboptimal performance.

First, we consider the case of binary pure state discrimination, where a locally-greedy (LG) technique is known to be optimal.  We restrict the LG algorithm to the same action space as the RLNN to remove any gap due to action space quantization, and compare the resulting success probabilities. Results are depicted in Fig.~\ref{fig:LG}, and indicate that the RLNN matches or almost matches the LG algorithm when $n=10$ but develops a performance gap for $n=20$.

\begin{figure}[h!]
\begin{minipage}{.45\textwidth}
\begin{center}
\begin{tikzpicture}[scale = 0.83]
\begin{axis}[
legend cell align={left},
legend style={fill opacity=0.8, draw opacity=1, text opacity=1, at={(0.97,0.03)}, anchor=south east, draw=white!80!black},
tick align=outside,
tick pos=left,
x grid style={white!69.0196078431373!black},
xlabel={trial},
xmin= 0.9, xmax=10.1,
xtick style={color=black},
y grid style={white!69.0196078431373!black},
ylabel={$\text{P}_{\text{succ}}$},
ymin=0.51, ymax=1,
ytick style={color=black}
]

\addplot [only marks, green!50.1960784313725!black, mark=square*, mark size=3, mark options={solid}]
table {%
1 0.561795317617062
2 0.57368418098813
3 0.655481018342062
4 0.717706669915716
5 0.725608266910164
6 0.773963448146067
7 0.831357656028497
8 0.953258265471435
9 0.979315760701043
10 0.987315043256477
};
\addlegendentry{LG}
\addplot [only marks, black, mark=triangle*, mark size=3, mark options={solid}]
table {%
1 0.554504605893847
2 0.567754206584423
3 0.643542449485155
4 0.704844387749806
5 0.716616716159488
6 0.752980918261999
7 0.817476107739502
8 0.939217804967306
9 0.976083292538998
10 0.984531804527923
};
\addlegendentry{RLNN}
\end{axis}

\end{tikzpicture}
  \caption{\label{fig:LG_n_10} Success probability for $m=2$, $n=10$ where all candidate states are pure. Success probability for the RLNN is based on 750 training iterations for each round.}
\end{center}
\end{minipage}
\begin{minipage}{.05\textwidth}

\hspace*{0.1cm}

\end{minipage}
\begin{minipage}{.45\textwidth}
\begin{center}
\begin{tikzpicture}[scale = 0.83]
\begin{axis}[
legend cell align={left},
legend style={fill opacity=0.8, draw opacity=1, text opacity=1, at={(0.97,0.03)}, anchor=south east, draw=white!80!black},
tick align=outside,
tick pos=left,
x grid style={white!69.0196078431373!black},
xlabel={trial},
xmin= 0.8, xmax=10.2,
xtick style={color=black},
y grid style={white!69.0196078431373!black},
ylabel={$\text{P}_{\text{succ}}$},
ymin=0.553631892, ymax=1.000564268,
ytick style={color=black}
]

\addplot [only marks, green!50.1960784313725!black, mark=square*, mark size=3, mark options={solid}]
table {%
1 0.58619
2 0.58705
3 0.656329
4 0.70635548
5 0.7254644
6 0.758215
7 0.84851124
8 0.86871842
9 0.92672568
10 0.98024916
};
\addlegendentry{LG}
\addplot [only marks,  black, mark=triangle*, mark size=3, mark options={solid}]
table {%
1 0.573947
2 0.5783
3 0.611368
4 0.68239611
5 0.70838644
6 0.728872254
7 0.82221578
8 0.84118215
9 0.88806859
10 0.9624956
};
\addlegendentry{RLNN}
\end{axis}

\end{tikzpicture}

  \caption{\label{fig:LG} Success probability for $m=2$, $n=20$ where all candidate states are pure. Success probability for the RLNN is based on 1000 training iterations for each round.  }
\end{center}
\end{minipage}
\end{figure}
Next, we consider the performance of multiple state discrimination where $m=3$. For $n<8$ we can compare directly to the collective SDP. For $n \geq 8$, computing $P_{\text{succ}}$ via SDP techniques is infeasibly slow, so we instead look at the RLNN training curve shape to determine whether the neural network converges to a steady solution.

As a first test of RLNN performance when $m=3$, we consider the case of pure states with $n=5$  and plot the success probability vs SDP as well as the training curves in Fig.~\ref{fig:n5_vs_SDP}.The RLNN success probability plateaus after approximately 100 training iterations, and the RLNN success probability comes close to the collective SDP in each case.  
\begin{figure}[h!]
\begin{minipage}{.45\textwidth}
\begin{center}
 \begin{tikzpicture}[scale = 0.83]

\begin{axis}[
legend cell align={left},
legend style={fill opacity=0.8, draw opacity=1, text opacity=1, at={(0.97,0.03)}, anchor=south east, draw=white!80!black},
tick align=outside,
tick pos=left,
x grid style={white!69.0196078431373!black},
xlabel={trial},
xmin=-0.45, xmax=9.45,
xtick style={color=black},
y grid style={white!69.0196078431373!black},
ylabel={$\text{P}_{\text{succ}}$},
ymin=0.358796286210958, ymax=0.982339287599099,
ytick style={color=black}
]
\addplot [only marks, green!50.1960784313725!black, mark=square*, mark size=2.5, mark options={solid}]
table {%
0 0.39312145241583
1 0.474592886740007
2 0.527887399675838
3 0.532525808233795
4 0.656214746622209
5 0.66783100528854
6 0.675707236641309
7 0.693125432600173
8 0.830779268421552
9 0.953996423899638
};
\addlegendentry{SDP}
\addplot [only marks, black, mark=*, mark size=2.5, mark options={solid}]
table {%
0 0.387139149910419
1 0.461933280111698
2 0.513039247515642
3 0.511169911773439
4 0.64187926306082
5 0.658289172298126
6 0.672161501012573
7 0.678862803771012
8 0.812248459809263
9 0.941300492360904
};
\addlegendentry{RLNN}
\end{axis}

\end{tikzpicture}

  \caption{\label{fig:n5_vs_SDP} 10 trials with $m=3$, $n=5$. Success probability for RLNN after 300 training iterations compared to success probability of SDP}
\end{center}
\end{minipage}
\begin{minipage}{.05\textwidth}

\hspace*{0.1cm}

\end{minipage}
\begin{minipage}{.45\textwidth}
%\begin{figure}
\begin{center}
 \begin{tikzpicture}[scale = 0.83]

\definecolor{color0}{rgb}{0.12156862745098,0.466666666666667,0.705882352941177}
\definecolor{color1}{rgb}{1,0.498039215686275,0.0549019607843137}
\definecolor{color2}{rgb}{0.172549019607843,0.627450980392157,0.172549019607843}
\definecolor{color3}{rgb}{0.83921568627451,0.152941176470588,0.156862745098039}
\definecolor{color4}{rgb}{0.580392156862745,0.403921568627451,0.741176470588235}
\definecolor{color5}{rgb}{0.549019607843137,0.337254901960784,0.294117647058824}
\definecolor{color6}{rgb}{0.890196078431372,0.466666666666667,0.76078431372549}
\definecolor{color7}{rgb}{0.737254901960784,0.741176470588235,0.133333333333333}
\definecolor{color8}{rgb}{0.0901960784313725,0.745098039215686,0.811764705882353}

\begin{axis}[
tick align=outside,
tick pos=left,
x grid style={white!69.0196078431373!black},
xlabel={training iteration},
xmin=-4, xmax=304,
xtick style={color=black},
y grid style={white!69.0196078431373!black},
ylabel={$\text{P}_{\text{h}}- \text{P}_{\text{RLNN}}$},
ymin=-0.0065636783445757, ymax=0.0780120714060898,
ytick style={color=black}
]
\addplot [line width=0.04pt, color0, mark=*, mark size=2, mark options={solid}]
table {%
10 0.0722363726323775
20 0.0433247773447371
30 0.0257064933842666
40 0.0235686996174049
50 0.0202365047028327
60 0.0204282644400116
70 0.0173913961617089
80 0.0161076066688466
90 0.0159920968716531
100 0.0150257567562453
110 0.0152545248579041
120 0.0149584236125119
130 0.0150288837312212
140 0.0146341326509365
150 0.0141401498904324
160 0.0147998856067789
170 0.0130901804186637
180 0.0131624326233393
190 0.0135341354564861
200 0.0124470719060934
210 0.0120114313469187
220 0.0129637386466959
230 0.0123663523840919
240 0.0129867981347063
250 0.0143552591560967
260 0.0121013876956948
270 0.0133098541901009
280 0.0120594606181913
290 0.0126959315387336
};
\addplot [line width=0.04pt, color1, mark=*, mark size=2, mark options={solid}]
table {%
10 0.0577087291000724
20 0.0443996523888072
30 0.0349782354301877
40 0.0300193993130994
50 0.0235513931083037
60 0.024298665409209
70 0.0247667555180831
80 0.0197705026938213
90 0.021296510101144
100 0.0214609596708776
110 0.0202374559290082
120 0.0179671856821028
130 0.0186442502689605
140 0.0195712207270454
150 0.0199187334129699
160 0.0183114306065438
170 0.0137955700865675
180 0.0124980643184501
190 0.0129686397630656
200 0.0134893913610948
210 0.0147817463090208
220 0.010969104912735
230 0.0112587665557866
240 0.0142595895868787
250 0.0144640097824124
260 0.0127607053199217
270 0.0107184353245026
280 0.0114943311827466
290 0.00954183299041411
};
\addplot [line width=0.04pt, color2, mark=*, mark size=2, mark options={solid}]
table {%
10 0.0615860626122139
20 0.0413364475818895
30 0.0346892141282296
40 0.0265269314356863
50 0.0285121838024762
60 0.0255282072266504
70 0.0238017589771568
80 0.0235282244182483
90 0.0256044927326742
100 0.0219832132946159
110 0.0224389586997733
120 0.0217295347584233
130 0.0242430893616249
140 0.0174945342704299
150 0.0192853256264743
160 0.0221167542833832
170 0.0177204421754774
180 0.0211678297925157
190 0.0140980753757782
200 0.016038933065869
210 0.00981079395302464
220 0.0099109313502197
230 0.0157287204190966
240 0.0153718418701601
250 0.0134027028846383
260 0.0163421984922186
270 0.0133586946226099
280 0.00964188158346524
290 0.0143354835613895
};
\addplot [line width=0.04pt, color3, mark=*, mark size=2, mark options={solid}]
table {%
10 0.0303119736908054
20 0.0159579915219424
30 0.0161796918520205
40 0.0091384431212278
50 0.01442053823408
60 0.00579985884598233
70 0.00673722543993771
80 0.00515702509769433
90 0.00460328524675579
100 0.0101244957449917
110 0.00833355171503258
120 0.0118932255134889
130 0.00575269171484438
140 0.00449455621014216
150 0.00105225225285399
160 0.0023591945615844
170 0.00173012657967297
180 0.00436662687699441
190 0.00261260458982759
200 0.00585954288725554
210 -0.00271932608318182
220 0.00301131281756339
230 0.0044945933710937
240 0.00336322087394669
250 -0.000438519692903538
260 0.00166183851866109
270 0.00788686652501702
280 0.00460879344119625
290 0.00354573562873628
};
\addplot [line width=0.04pt, color4, mark=*, mark size=2, mark options={solid}]
table {%
10 0.0741677191446959
20 0.047896063461782
30 0.0397140139400544
40 0.0320586747901365
50 0.0285405881362748
60 0.0271684332078491
70 0.0257802193255501
80 0.0245363601727379
90 0.0225607999538286
100 0.0248661767763322
110 0.0217676696541164
120 0.0227720164614033
130 0.0237531860517218
140 0.0188531550148043
150 0.0210911915261756
160 0.0183854001858209
170 0.0185001254627215
180 0.022130976026023
190 0.0213114752615001
200 0.0213179038077987
210 0.0209978710221627
220 0.0217982478967444
230 0.0188841996627893
240 0.0199950774346183
250 0.0195054096889443
260 0.0207896460451139
270 0.0173699173203233
280 0.0219593799014239
290 0.0185308086122891
};
\addplot [line width=0.04pt, color5, mark=*, mark size=2, mark options={solid}]
table {%
10 0.0224646647582132
20 0.0128008789010414
30 0.00926588907220738
40 0.00851225717945214
50 0.00835110486290325
60 0.00768705506877387
70 0.00742200567124879
80 0.00632841774038667
90 0.006964183345046
100 0.00639627483708499
110 0.0061194034363819
120 0.00649535467655843
130 0.00717812729259487
140 0.00608629112458253
150 0.00615295693919987
160 0.00632615210213722
170 0.00633655809142009
180 0.00622912252663915
190 0.0059618406106276
200 0.00599098292507555
210 0.00594861271790947
220 0.00658486388093915
230 0.00578234619575851
240 0.00657987128977283
250 0.00580267347031083
260 0.00646395868209726
270 0.00543339907553081
280 0.00509433163058687
290 0.00598230250541054
};
\addplot [line width=0.04pt, color6, mark=*, mark size=2, mark options={solid}]
table {%
10 0.0447326131260619
20 0.0311215081512662
30 0.0265309818642432
40 0.0209455825506307
50 0.0199649750156692
60 0.0215752323262486
70 0.017087155307739
80 0.01863479523486
90 0.0194729262849141
100 0.0196568596366259
110 0.0174030502174116
120 0.0205819244615402
130 0.0175951494072571
140 0.0169872433110632
150 0.0153067039345434
160 0.0146216289275729
170 0.0193006642435357
180 0.0179191355135441
190 0.0190906637948637
200 0.0165812133095807
210 0.0180954107299875
220 0.0182574564114016
230 0.0175041677706192
240 0.0141256524519396
250 0.0140938887466174
260 0.0147394536327491
270 0.013436323078026
280 0.0110896314120535
290 0.0148481521601962
};
\addplot [line width=0.04pt, white!49.8039215686275!black, mark=*, mark size=2, mark options={solid}]
table {%
10 0.063143492382847
20 0.0405624481595563
30 0.0322857629042603
40 0.0314233561567124
50 0.030956275717261
60 0.0288525083377904
70 0.0257862534385658
80 0.0276531943922702
90 0.0226114032647808
100 0.0245769138502068
110 0.0239923901355301
120 0.0249442253064526
130 0.0240727518502978
140 0.0255981258036118
150 0.0215505421342523
160 0.0226570704715627
170 0.0207113906787098
180 0.0173453077209036
190 0.017589428856099
200 0.0175476120312523
210 0.016306794985758
220 0.012194154390389
230 0.0151702570734299
240 0.0142309772636611
250 0.0139451512924456
260 0.0176280260542749
270 0.0170701558926933
280 0.0153001381346373
290 0.0142626288291603
};
\addplot [line width=0.04pt, color7, mark=*, mark size=2, mark options={solid}]
table {%
10 0.0357072968904243
20 0.0255419962696447
30 0.0200159727951113
40 0.0197715611948346
50 0.0178991683211629
60 0.0178347091202971
70 0.0156136420481734
80 0.0167613517677304
90 0.0162514307924142
100 0.0151464250206024
110 0.0136470755829894
120 0.0165731748576244
130 0.0146745645329647
140 0.0134020069346232
150 0.0143202379170163
160 0.0133518070574579
170 0.0139080089661286
180 0.0149036620475788
190 0.0151883465411132
200 0.014736634534011
210 0.0144647719858305
220 0.0119522625677861
230 0.0142433620219283
240 0.0137086469948443
250 0.0133082028118644
260 0.0128140441574953
270 0.0123454642295283
280 0.0153778095376141
290 0.0126596066283085
};
\addplot [line width=0.04pt, color8, mark=*, mark size=2, mark options={solid}]
table {%
10 0.0483901298149503
20 0.0339131753525085
30 0.0296741325869155
40 0.026200286378303
50 0.0245160884138691
60 0.021810445088228
70 0.023517731592669
80 0.0255168380017984
90 0.0231486362454166
100 0.0241138561691251
110 0.0209729954659003
120 0.020062253566538
130 0.02434112660796
140 0.0212355955081821
150 0.021059315145653
160 0.0230823782375396
170 0.0206602216144894
180 0.0215018679704109
190 0.0199889069081762
200 0.020674181391844
210 0.019002470416166
220 0.022508054417899
230 0.0213291643048314
240 0.022103356474369
250 0.0206533510391097
260 0.0224843762931686
270 0.0212341063375712
280 0.0236441071331713
290 0.0213558964603564
};
\end{axis}

\end{tikzpicture}

  \caption{\label{fig:n5_training} Difference between RLNN reward and SDP success probability as a function of training iteration. We observe that the RLNN success probability approximately plateaus after 100 training iterations}
\end{center}
\end{minipage}
\end{figure}

We then examine the training curves for general state discrimination when $m=3$ as a function of $n$, with results shown in Fig.~\ref{fig:n20}. Although stable plateaus are reached for both $n=10$ and $n=20$, as $n$ increases the RLNN spends more training iterations learning not to re-measure subsystems. In the case where $n=50$, approximately 500 training iterations are spent learning not to re-measure subsystems, leading to a negative initial reward. This leads to the question of whether it is possible to extend the RLNN performance to a larger number of subsystems by predetermining a close-to-optimal ordering, 

\begin{figure}[h!]
%\begin{figure}
\includegraphics[scale = 0.48]{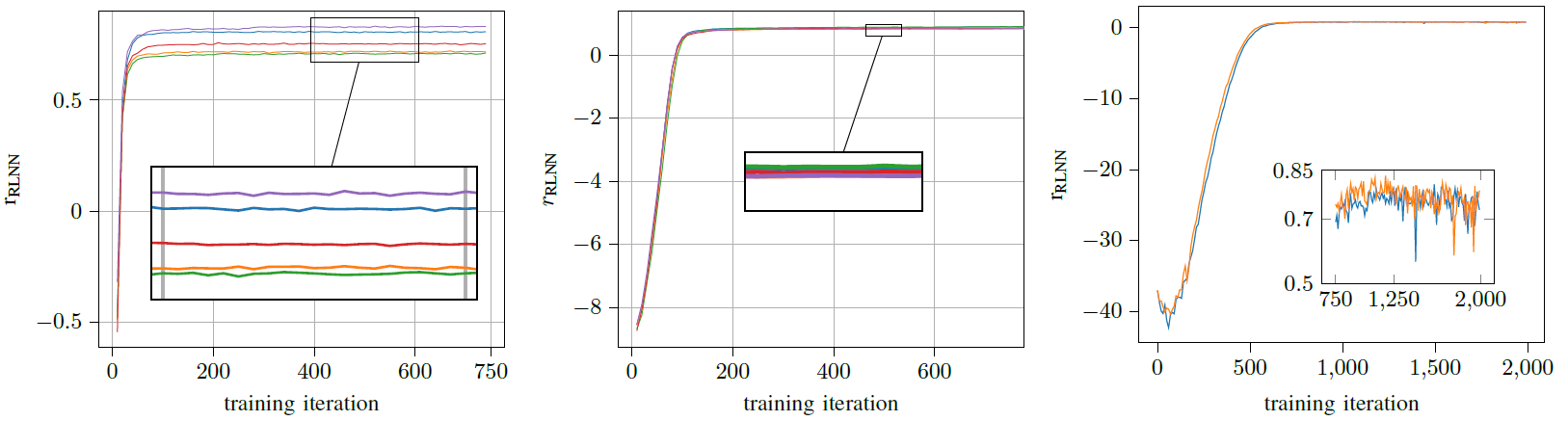}
  \caption{\label{fig:n20} Training curves for five independent trials where $n=10$ (right), $n=20$ (centre), and one trial of $n=50$ (left). As $n$ increases, the training curve becomes less stable, and the number of iterations required for the reward to reach its maximal value increases. For $n=50$, the shape of the training curve changes to include an initial dip in reward before convergence to the ideal and we also observe less stability.}
\end{figure}

\section{Robustness under noise}
\label{sec:noise}

We demonstrate that the success probability is stable when the candidate states are subject to a small perturbation. Consider an over-rotation noise model where the perturbation is parametrised by rotation angle $\theta$ with

\begin{align*}
  \tilde{\rho}_{j}^{(k)}(\theta) = U(\theta) \rho_{j}^{(k)} U^{\dag}(\theta),
\end{align*}
where  $ \tilde{\rho}_{j}^{(k)}(\theta)$ is the noisy state and we set the rotation matrix as $
    U(\theta) = \begin{psmallmatrix}
    \cos(\theta) & \sin(\theta) \\
    - \sin(\theta) & \cos(\theta)
    \end{psmallmatrix}$. 
We prove that the error due to noise is negligible for sufficiently small enough perturbations, indicating that the RLNN adaptive method is still close-to-optimal when the candidate states are subjected to sufficiently small amounts of unitary noise. 

\begin{thm}
Consider candidate set $\{\rho_{j}\}|_{j=1}^{m}$ with prior $\mathbf{q}$. Denote by $\text{P}_{\text{succ}}\Big( \{\rho_{j}\} \Big)$ the probability of success using the optimal locally adaptive method on the original state set. Likewise, let $\text{P}_{\text{succ}}\Big( \{\tilde{\rho}_{j}(\theta)\} \Big)$ be the success probability for the noisy state set. Then for all $\theta$,
\begin{align*}
    \left| \text{P}_{\text{succ}}\Big( \{\rho_{j}\} \Big) - \text{P}_{\text{succ}}\Big( \{\tilde{\rho}_{j}(\theta)\} \Big) \right| \leq 4n \sin\left(\frac{|\theta|}{2}\right),
\end{align*}
where $n$ is the number of subsystems.
\end{thm}
\noindent{\bf Proof:} See Appendix B for a complete proof.

Finally, we generate five candidate state sets with $m=3$, $d=2$. For each candidate state set $\{\rho_{j}\}$, we train the neural network to find the optimal locally adaptive method. The original adaptive measurement scheme is then applied to the rotated state set $\{\tilde{\rho}_{j}(\theta)\}$, and we plot the gap in success probabilities $\text{diff}(\theta) \triangleq  \text{P}_{\text{succ}}\Big( \{\rho_{j}\} \Big) - \text{P}_{\text{succ}}\Big( \{\tilde{\rho}_{j}(\theta)\} \Big)$.
\begin{figure}[h!]
\begin{center}
 \begin{tikzpicture}[scale = 0.85]

\definecolor{color0}{rgb}{0.12156862745098,0.466666666666667,0.705882352941177}
\definecolor{color1}{rgb}{1,0.498039215686275,0.0549019607843137}
\definecolor{color2}{rgb}{0.172549019607843,0.627450980392157,0.172549019607843}
\definecolor{color3}{rgb}{0.83921568627451,0.152941176470588,0.156862745098039}
\definecolor{color4}{rgb}{0.580392156862745,0.403921568627451,0.741176470588235}

\begin{axis}[
log basis x={10},
tick align=outside,
tick pos=left,
x grid style={white!69.0196078431373!black},
xmin=0.000794328234724281, xmax=0.125892541179417,
ylabel = {$\text{diff}(\theta)$},
xlabel = {$\theta$},
xmode=log,
xtick style={color=black},
y grid style={white!69.0196078431373!black},
ymin=-0.00355691897097351, ymax=0.136489597521286,
ytick style={color=black}
]
\addplot [semithick, color0, mark=square*, mark size=1.5]
table {%
0 0
0.001 0.00256796147872929
0.005 0.00141099453930271
0.01 0.00238209129014016
0.05 0.00813755033981239
0.075 0.039782342200606
0.1 0.0659930222781959
};
\addplot [semithick, color1, mark=square*, mark size=1.5]
table {%
0 0
0.001 -0.00157839116105385
0.005 -0.00196389549405263
0.01 -0.000491996072550571
0.05 0.00931824755832167
0.075 0.0350194726598974
0.1 0.0631252655064621
};
\addplot [semithick, color2, mark=square*, mark size=1.5]
table {%
0 0
0.001 -0.00084577051124568
0.005 -0.00106537429081999
0.01 0.0002749204564505
0.05 0.00745042035534166
0.075 0.0330059185104812
0.1 0.0649731443393032
};
\addplot [semithick, color3, mark=square*, mark size=1.5]
table {%
0 0
0.001 0.000379552357007706
0.005 -0.00140302240639567
0.01 0.000973194153680912
0.05 0.0113069900352779
0.075 0.0483336240463259
0.1 0.080869877515277
};
\addplot [semithick, color4, mark=square*, mark size=1.5]
table {%
0 0
0.001 0.00352098995022798
0.005 0.00260526726230781
0.01 0.00419652143885019
0.05 0.0206993219781264
0.075 0.0659993099492657
0.1 0.129896574044365
};
\end{axis}

\end{tikzpicture}

  \caption{\label{fig:advs_n2_m2} Gap in success probability as a function of rotation parameter $\theta$ for five trials where $m=n=3$. We observe that the gap is negligible for sufficiently small $\theta$.}
\end{center}
\end{figure}
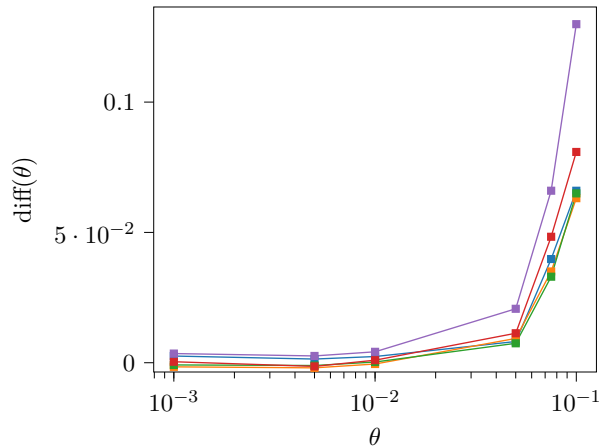

\section{Conclusion}
\label{sec:conclusion}

We apply RLNN to calculate a near-optimal locally-adaptive measurement scheme for multiple state discrimination. We provide preliminary results for the neural network performance in cases where the locally-adaptive probability of success is known, and show that the network can achieve good performance when the total number of subsystems is 10 or fewer. This performance holds even when the candidate states are subjected to small perturbations. For cases where the exact locally optimal protocol is not known, we compare the RLNN performance to an SDP upper bound and find that for each trial the RLNN comes close to the upper bound. Additionally, we introduce a min-entropy based locally adaptive approach which reduces to the optimal local approach for binary pure states, and show that the RLNN meets or exceeds this approach in every trial. 

Finally, we characterize types of candidate state sets where there is a gap between optimal locally adaptive algorithms and optimal collective algorithms. While previous work has demonstrated gaps for candidate state sets with a more complex structure, we provide state sets where a gap exists for the simplest possible case of binary state discrimination with three depolarized qubits, as well as a binary state discrimination problem with an even smaller system composed of two qutrits. Future work aims to extend the RLNN performance to a larger number of subsystems, as well as to characterize the maximal gap between the optimal local and optimal collective success probability as a function of the number of candidate states and subsystems.

\section*{Competing Interests}
The authors declare that there are no competing interests.

\section*{Acknowledgments}
The authors would like to thank Narayanan Rengaswamy and Dhruva Sambrani for helpful discussions. The work of Brandsen and Pfister was supported in part by the National Science Foundation (NSF) under Grant No. 1908730 and 1910571. Stubbs was supported in part by a National Science Foundation Graduate Research Fellowship under Grant No. DGE-1644868. Any opinions, findings, conclusions, and recommendations expressed in this material are those of the authors and do not necessarily reflect the views of these sponsors.
% \IEEEtriggeratref{10}

\balance

\bibliographystyle{ieeetr}

\appendix 

\section{Proof of Theorem 1}

The ``only if'' direction of the proof follows immediately from~\cite{Virmani_2001}, where it was shown that locally adaptive methods are optimal for any binary pure state discrimination problem (including discrimination problems involving entangled states.) \\~\\
We now demonstrate that even in the case of binary state discrimination, purity is a necessary condition for optimal state discrimination for \emph{any} number of subsystems $m$. We consider the following set of two qutrit candidate states:
\begin{align*}
    \rho_{+} & \triangleq \Big( \frac{1}{2} \ket{0}\!\!\bra{0} + \frac{1}{2} \ket{1}\!\!\bra{1} \Big) \otimes \Big( \frac{1}{2} \ket{0}\!\!\bra{0} + \frac{1}{2} \ket{1}\!\!\bra{1} \Big) \\
    \rho_{-} & \triangleq \frac{1}{3} \Big( \sum_{j=0}^{2} \ket{j}\otimes \ket{j} \Big) \Big( \sum_{k=0}^{2} \bra{k}\otimes \bra{k} \Big).
\end{align*}

To our knowledge, this is the smallest dimensional system where binary state discrimination cannot be optimally performed via projective locally adaptive measurements.

\begin{figure}[h!]
\begin{center}
%\begin{figure}
\includegraphics[scale = 0.6]{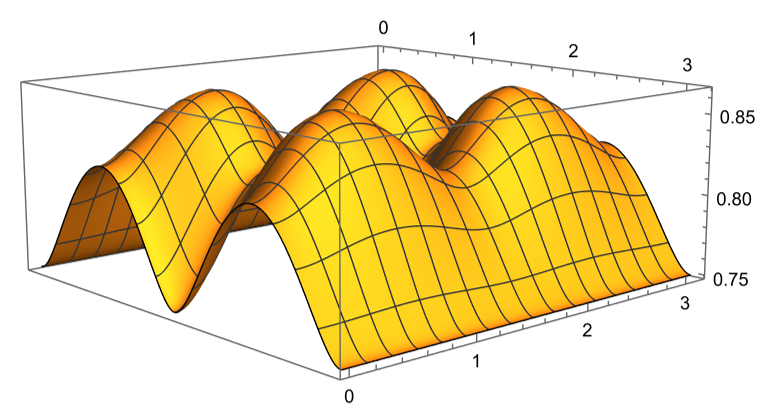}
  \caption{\label{fig:n20} $P_{\text{succ}}(\theta, \omega)$ plotted over the range $0 \leq \theta \leq \pi$ and $0 \leq \omega \leq \pi$.}
  \label{fig:RLNN3Dfig}
  \end{center}
\end{figure}

The collective success probability is $P_{\text{succ}} = \frac{1}{48} (33 + \sqrt{129}) \approx 0.924121$. We now utilize a computer-assisted proof to demonstrate that the gap between the collective and local success probability is strictly positive. To this aim, we introduce a set of closely quantized measurements, find the optimal measurement strategy from the quantized set, and then demonstrate that the success probability is smooth enough that \emph{any} projective local measurement strategy must be strictly worse than the optimal collective measurement.  \\~\\
The set of projective qutrit measurements with parameters $\phi$, $\theta$, and $\omega$ is given by:
\begin{align*}
    \Big\{ \ket{u_{1}(\phi, \theta, \omega)}\!\!\bra{u_{1}(\phi, \theta, \omega)}, \ket{u_{2}(\phi, \theta, \omega)}\!\!\bra{u_{2}(\phi, \theta, \omega)}, \ket{u_{3}(\phi, \theta, \omega)}\!\!\bra{u_{3}(\phi, \theta, \omega)} \Big\}
\end{align*}
where
\begin{align*}
 \ket{u_{1}(\phi, \theta, \omega)} & \triangleq \begin{pmatrix}
 -\cos(\omega)\sin(\phi) + \cos(\phi)\cos(\theta)\sin(\omega) \\
 \cos(\phi) \cos(\omega) + \cos(\theta) \sin(\phi) \sin(\omega) \\
 -\sin(\theta) \sin(\omega)
 \end{pmatrix} \\
  \ket{u_{2}(\phi, \theta, \omega)} & \triangleq \begin{pmatrix}
 \cos(\phi)\cos(\theta)\cos(\omega) + \sin(\phi)\sin(\omega) \\
 \cos(\theta) \cos(\omega) \sin(\phi) - \cos(\phi) \sin(\omega) \\
 - \cos(\omega) \sin(\theta)
 \end{pmatrix} \\
 \ket{u_{3}(\phi, \theta, \omega)} & \triangleq \begin{pmatrix}
 \cos(\phi) \sin(\theta) \\
 \sin(\phi) \sin(\theta) \\
 \cos(\theta)
 \end{pmatrix}. \\
\end{align*}
Any locally adaptive strategy will then consist of a sequence of measurements, such that $\hat{\Pi}(\phi, \theta, \omega)$ is implemented on the first subsystem and, given measurement outcome $d_{1}$ from the first measurement, $\hat{\Pi'}(d_{1}) = \{\Pi'_{+}(d_{1}), \Pi'_{-}(d_{1})\}$ is implemented on the second subsystem. Evidently, the optimal measurement on the second subsystem will always be the Helstrom measurement given the updated prior and candidate states. \\
Thus, finding the optimal locally adaptive strategy is equivalent to finding the optimal first measurement $\hat{\Pi}(\phi, \theta, \omega)$. 
% Define $\text{P}_{\text{s}}(\phi, \theta, \omega)$ as
% \begin{align*}
%     \text{P}_{\text{s}}(\phi, \theta, \omega) = \sum_{x \in \{+, -\}} \sum_{j =1}^{3} \text{Pr}(\rho = \rho_{x}) \text{Tr}\Bigg[\Big( \Pi_{j}(\phi, \theta, \omega) \otimes \mathbb{I} \Big) \rho_{x}\Bigg] \text{Tr}\Bigg[\Big( \mathbb{I} \otimes \Pi'_{x}(d = j) \Big) \rho_{x}(d=j)\Bigg]
% \end{align*}
% where $\rho_{x}(d=j)$ is the updated state of $\rho_{x}$ given the previous measurement outcome corresponds to element $\Pi_{j}(\phi, \theta, \omega)$. 
Using the state set above with starting prior $q = \frac{1}{2}$, the success probability is given by
{\small
\begin{align*}
     \text{P}_{\text{s}}(\phi, \theta, \omega) &= \sum_{j =1}^{3} \text{Tr}\Big[\Big( \Pi_{j}(\phi, \theta, \omega) \otimes \mathbb{I} \Big) \Big(\frac{1}{2} \rho_{+} + \frac{1}{2} \rho_{-} \Big)\Big]  \Bigg( \frac{1}{2} + \frac{1}{2} \Big \| \text{Pr}(\rho = \rho_{+} \big| \phi, \theta, \omega, d_{1} = j) \rho_{+}(d_{1} = j)\\
    &- \Big(1- \text{Pr}(\rho = \rho_{+} \big| \phi, \theta, \omega, d_{1} = j)\Big) \rho_{-}(d_{1} = j) \Big\|_{1} \Bigg),
    \end{align*}}
    where $\rho_{\pm}(d_{1} = j)$ is the post-measurement state for $\rho_{\pm}$ given that the measurement outcome $d_{1}$ is observed to correspond to $\Pi_{j}(\phi, \theta, \omega)$ on the first subsystem. We now simplify the above expression so that it can be computed directly in Mathematica. First, we note that 
    \begin{align*}
        \rho_{+}(d_{1} = j) & \triangleq \ket{u_{j}(\phi, \theta, \omega)}\!\!\bra{u_{j}(\phi, \theta, \omega)} \otimes \Big( \frac{1}{2}\ket{0}\!\!\bra{0} + \frac{1}{2}\ket{1}\!\!\bra{1} \Big) \\
        \rho_{-}(d_{1} = j) & \triangleq \ket{u_{j}(\phi, \theta, \omega)}\!\!\bra{u_{j}(\phi, \theta, \omega)} \otimes \ket{u_{j}(\phi, \theta, \omega)}\!\!\bra{u_{j}(\phi, \theta, \omega)}
    \end{align*}
    and likewise 
    \begin{align*}
    \text{Tr}\Big[\Big( \Pi_{j}(\phi, \theta, \omega) \otimes \mathbb{I} \Big) \Big(\frac{1}{2} \rho_{+} + \frac{1}{2} \rho_{-} \Big)\Big] &= \frac{1}{2} \Bigg( \frac{1}{3} +  \text{Tr}\Big[\Big( \Pi_{j}(\phi, \theta, \omega) \otimes \mathbb{I} \Big)  \rho_{+} \Big] \Bigg)  \\
    &= \frac{1}{2} \Bigg( \frac{1}{3} +  \text{Tr}\Big[ \Pi_{j}(\phi, \theta, \omega) \Big( \frac{1}{2}\ket{0}\!\!\bra{0} + \frac{1}{2}\ket{1}\!\!\bra{1} \Big) \Big] \Bigg), 
    \end{align*}
  where the first line follows from noting that all measurement outcomes are equally likely on the maximally entangled state. Finally, we can rewrite the success probability as

\begin{align*}
     \text{P}_{\text{s}}(\phi, \theta, \omega) &= \sum_{j =1}^{3} \frac{1}{2} \Bigg( \frac{1}{3} +  \text{Tr}\Big[ \Pi_{j}(\phi, \theta, \omega) \frac{\ket{0}\!\!\bra{0} + \ket{1}\!\!\bra{1}}{2} \Big] \Bigg)   \Bigg( \frac{1}{2} + \frac{1}{2} \Big \|\text{Pr}(\rho = \rho_{+} \big| \phi, \theta, \omega, d_{1} = j)\\
     & \times \frac{\ket{0}\!\!\bra{0} + \ket{1}\!\!\bra{1}}{2}- \text{Pr}(\rho = \rho_{-} \big| \phi, \theta, \omega, d_{1} = j) \ket{u_{j}(\phi, \theta, \omega)}\!\!\bra{u_{j}(\phi, \theta, \omega)} \Big\|_{1} \Bigg) \\
    &= \frac{1}{2} + \sum_{j=1}^{9} g_{k}(\phi, \theta, \omega) \big| f_{k}(\phi, \theta, \omega) \big|,
    \end{align*}
where in the last step we rewrite the success probability in terms of functions $g_{k}(\phi, \theta, \omega)$, which correspond to the probability of observing measurement outcome $\lfloor \frac{k}{3} \rfloor$ when measuring the first subsystem, and $f_{k}(\phi, \theta, \omega)$ which correspond to eigenvalues arising from the trace norm. More specifically, for a given $j \in \{1,2,3\}$, let $\lambda_{j,1}, \lambda_{j,2}, \lambda_{j,3}$ denote the three eigenvalues of the operator
\begin{align*}
    \text{Pr}(\rho = \rho_{+} \big| \phi, \theta, \omega, d_{1} = j) \frac{\ket{0}\!\!\bra{0} + \ket{1}\!\!\bra{1}}{2}- \text{Pr}(\rho = \rho_{-} \big| \phi, \theta, \omega, d_{1} = j) \ket{u_{j}(\phi, \theta, \omega)}\!\!\bra{u_{j}(\phi, \theta, \omega)}.
\end{align*}
The functions $\{ f_{k}(\phi, \theta, \omega) \}_{k=1}^9$ are then defined as $f_{3(j-1)+\ell}(\phi, \theta, \omega) := \lambda_{j,\ell}$ where $j, \ell \in \{ 1, 2, 3 \}$. 
\normalsize
Likewise, 
\begin{align*}
    g_{k}(\phi, \theta, \omega) &= 
    \frac{1}{4} \Big( \frac{1}{3} +  \text{Tr}\Big[ \Pi_{\lfloor \frac{k}{3} \rfloor}(\phi, \theta, \omega) \frac{\ket{0}\!\!\bra{0} + \ket{1}\!\!\bra{1}}{2} \Big] \Big) 
\end{align*}
    The full expression for $P_{s}(\phi, \theta, \omega)$ is independent of $\phi$ and therefore can be denoted as $P_{s}(\theta, \omega)$. While $P_{\text{s}}(\phi, \theta, \omega)$ can be computed directly via Mathematica, the closed form expression is extremely lengthy.  For completeness, we provide the code used to generate and plot $P_{s}(\phi, \theta, \omega)$ at https://github.com/SarahBrandsen/RLNN-QSD. In Figure~\ref{fig:RLNN3Dfig}, we plot $P_{s}(\theta, \omega)$ to show the continuity of the function and demonstrate graphically that $P_{s}(\theta, \omega)$ is upper bounded by 0.87.

To prove this, we quantize $\theta$ and $\omega$ into $10,000$ discrete values such that $\theta, \omega \in \{\frac{2\pi j}{10000}\}|_{j=1}^{10000}$, and find that the best success probability achieved is
\begin{align*}
    \frac{1}{384} \left(240+\sqrt{1054-42 \sqrt{5}}+\sqrt{2302+630 \sqrt{5}}\right) \approx 0.864325
\end{align*}
Then, we demonstrate that any error due to quantization is sufficiently small (i.e., the gap between the local and collective measurement strategy cannot be due to quantization.) Consider a fixed $\theta$ and $\omega$ and for any $\theta'$ and $\omega'$ satisfying $|\theta'-\theta|\leq \epsilon$ and $|\omega - \omega'| \leq \epsilon$, then

\begin{align*}
\Big| & P_{s}(\theta, \omega) - P_{s}(\theta', \omega') \Big|  \\
& = \Big| \sum_{j} g_{j}(\theta, \omega) \big|f_{j}(\theta, \omega) \big| - g_{j}(\theta', \omega') \big| f_{j}(\theta', \omega') \big| \Big| \\
& = \Big| \sum_{j} \Big(g_{j}(\theta, \omega) - g_{j}(\theta', \omega') + g_{j}(\theta', \omega')\Big) \big|f_{j}(\theta, \omega) \big| - g_{j}(\theta', \omega') \big| f_{j}(\theta', \omega') \big| \Big| \\
& \leq \sum_{j} \big|g_{j}(\theta, \omega) - g_{j}(\theta', \omega') \big| \big|f_{j}(\theta, \omega) \big| + \big|g_{j}(\theta', \omega')\big| \big| f_{j}(\theta, \omega) - f_{j}(\theta', \omega') \big| \\
& \leq  \left( \max_{j} \big|g_{j}(\theta, \omega) - g_{j}(\theta', \omega') \big| \right) \sum_{j} \big|f_{j}(\theta, \omega) \big| \\
& \hspace{10ex} + \left( \max_{j}  \big| f_{j}(\theta, \omega) - f_{j}(\theta', \omega') \big| \right) \sum_{j} \big|g_{j}(\theta', \omega')\big| 
\end{align*}
By mean-value theorem, there exists a point $(\theta_{r_j}, \omega_{r_j})$ along the line connecting $(\theta, \omega)$ and $(\theta', \omega')$ so that
\begin{align*}
|g_{j}(\theta, \omega) - g_{j}(\theta', \omega')| & = |\nabla g_{j}(\theta_{r_j}, \omega_{r_j}) \cdot (\theta - \theta', \omega - \omega')| \\[1ex]
& \leq \epsilon \left( \Big|\frac{\partial g_{j}}{\partial \theta} (\theta_{r_j}, \omega_{r_j})\Big| + \Big|\frac{\partial g_{j}}{\partial \omega} (\theta_{r_j}, \omega_{r_j})\Big| \right) 
\end{align*}
where the last line follows from Holder's inequality. Similarly, there exists a point $(\theta_{r'_j}, \omega_{r'_j})$ so that
\[
\big| f_{j}(\theta, \omega) - f_{j}(\theta', \omega') \big| \leq \epsilon \left( \Big|\frac{\partial f_{j}}{\partial \theta} (\theta_{r'_j}, \omega_{r'_j})\Big| + \Big|\frac{\partial f_{j}}{\partial \omega} (\theta_{r'_j}, \omega_{r'_j})\Big| \right) 
\]
Using the explicit formula for $g_{j}(\theta,\omega)$ and $f_{j}(\theta,\omega)$ we computed with Mathematica, it can be checked that
\begin{align}
& \max_{\theta,\omega} \sum_{j} \big|f_{j}(\theta, \omega)\big| \leq 9
& \max_{j,\theta,\omega} \left( \Big|\frac{\partial f_{j}}{\partial \theta} (\theta, \omega)\Big| + \Big|\frac{\partial f_{j}}{\partial \omega} (\theta, \omega)\Big| \right) \leq 12 \\[1ex]
& \max_{\theta,\omega} \sum_{j} \big|g_{j}(\theta, \omega)\big| \leq 3
& \max_{j,\theta,\omega} \left( \Big|\frac{\partial g_{j}}{\partial \theta} (\theta, \omega)\Big| + \Big|\frac{\partial g_{j}}{\partial \omega} (\theta, \omega)\Big| \right) \leq \frac{1}{4}
\end{align}
Hence, we finally conclude that \footnote{While the current proof shows an analytical bound of $\frac{153}{4}$, computer-assisted proofs may potentially show tighter bounds.}
\begin{align*}
\max_{\theta', \omega'} \Big|P_{s}(\theta, \omega) - P_{s}(\theta', \omega')\Big| & \leq \frac{1}{4} \epsilon \times 9 + 12 \epsilon \times 3 \\
&= \frac{153}{4} \epsilon
\end{align*}
Setting $\epsilon = \frac{\Pi}{10 000}$ then gives 
\[
\max_{\theta', \omega'} \Big|P_{s}(\theta, \omega) - P_{s}(\theta', \omega')\Big| \leq 0.015.
\]
It follows that the observed gap between the locally adaptive and collective measurement scheme (with an approximate magnitude of 0.06) persists even after accounting for quantisation.

\noindent We likewise demonstrate that for the candidate states
\begin{align*}
    \rho_{+} & \triangleq \begin{pmatrix}
    0.85009903 & 0.1343714 \\
    0.1343714 & 0.14990097
    \end{pmatrix}^{\otimes 3}\\
    \rho_{-} & \triangleq \begin{pmatrix}
    0.58134943 & 0.36607003 \\
    0.36607003 & 0.41865057
    \end{pmatrix}^{\otimes 3}
\end{align*}
with $q = \frac{1}{2}$, then $P_{\text{coll}} - P_{\text{local}} \geq 0.011$ even when the quantization for the qubit projective measurements is set to $10,000$. (It follows from the proof of Theorem 2, discussed in Appendix B, that the largest error due to quantization in this case would be upper bounded by $3 \times 2^{3} \times \frac{\pi}{10000}$ and thus is negligible.)

Finally, we utilize the result found in~\cite{Croke2017} to demonstrate that even for pure state state discrimination, if $n \geq 3$, then for any $m$ there exists a candidate state with a gap. We restate the result found in~\cite{Croke2017} and provide a simplified proof which draws on their later work in~\cite{weir2018optimal}.  Consider as an example the case where $n=2$, $m=3$ and where each candidate state set consists of the trine ensemble, defined to be symmetric with 
\begin{align*}
    \rho_{j} &\triangleq \Big(U^{j} \ket{0}\!\!\bra{0} (U^{j})^{\dag}\Big)^{\otimes 2} \\
    &= (U \otimes U)^{j} \ket{00}\bra{00} \Big((U \otimes U)^{j}\Big)^{\dag},
\end{align*}
where $U \triangleq \begin{psmallmatrix}
    \cos(\frac{2\pi}{3}) & -\sin(\frac{2\pi}{3}) \\
    \sin(\frac{2\pi}{3}) & \cos(\frac{2\pi}{3})
    \end{psmallmatrix}$
and $\bm{q} = [1/3, 1/3, 1/3]$. 
Since $(U \otimes U)^{m} = \mathbb{I}$, and the starting prior is balanced, the PGM is optimal~\cite{Ban_SymmetricPGM}, with a corresponding collective success probability of $P_{\text{coll}} = \frac{1}{6}(3 + 2\sqrt{2}) \approx 0.971$. 

We now demonstrate that the optimal local strategy is to measure the first subsystem with an ``anti-trine'' measurement, defined as $\hat{\Pi}_{\text{AT}}=  \{ \frac{2}{3} U^{\frac{1}{2}}\rho_{j}(U^{\frac{1}{2}})^{\dag} \}|_{j=1}^{3}$, such that each measurement outcome is orthogonal to one of the candidate states. After obtaining the measurement outcome for the first subsystem, the updated prior is a permutation of $\mathbf{q} = [\frac{1}{2}, \frac{1}{2}, 0]$, and the second subsystem is measured according to the optimal measurement for the remaining two candidate states. \\
The most general local approach is to implement measurement $\hat{\Pi}_{1} = \big\{ \Pi_{1, j} \big\}|_{j=1}^{m}$ on the first subsystem. We may label the result of the first subsystem $\text{out}_{1}$. Then the second and last measurement can be chosen as $\hat{\Pi}_{2}(\text{out}_{1}) = \big\{ \Pi_{2, j}(\text{out}_{1}) \big\}|_{j=1}^{3}$ and is in general allowed to depend on the outcome $\text{out}_{1}$. It is conventional to label the elements of the second measurement such that state $\rho$ is decoded as $\rho_{j}$ if measurement element $j$ is obtained in the final round. Then

\begin{align*}
   \text{P}_{\text{succ}}\Big( \hat{\Pi}_{1}, \big\{ \hat{\Pi}_{2}(d_{1}) \big\}  \Big) &= \sum_{d_{1} = 1}^{m} \text{Pr}[\text{out}_{1} = \Pi_{d_{1}}^{(1)}] \text{P}_{\text{succ}}\big(\hat{\Pi}^{(2)}(d_{1}) \Big| \text{out}_{1} = \Pi_{d_{1}}^{(1)}  \big) \\
   & \leq \max_{ \hat{\Pi}_{1}, \big\{ \hat{\Pi}_{2}(d_{1}) \big\}}  \text{P}_{\text{succ}} \Big( \hat{\Pi}^{(2)}(d_{1})  \Big| \text{out}_{1} = \Pi_{d_{1}}^{(1)}  \Big)
\end{align*}
Thus, the second line presents an upper bound on the success probability of \emph{any} locally adaptive strategy, as it gives the success probability of the best possible measurement sequence. 

It follows that a sufficient condition for the optimality of the anti-trine based method is 
\begin{align*}
\text{P}_{\text{succ}}\Big( \hat{\Pi}_{\text{AT}}, \big\{ \hat{\Pi}_{2}^{*}(d_{1}) \big\}  \Big) & \geq \max_{ \hat{\Pi}_{1}, \big\{ \hat{\Pi}_{2}(d_{1}) \big\}}  \text{P}_{\text{succ}} \Big( \hat{\Pi}^{(2)}(d_{1})  \Big| \text{out}_{1} = \Pi_{d_{1}}^{(1)}  \Big)
\end{align*}

By symmetry, we know the expected success probability for the anti-trine is equivalent regardless of which outcome is obtained, and can immediately compute $
   \text{P}_{\text{succ}}(\hat{\Pi}_{\text{AT}}) =  0.933$.
From~\cite{weir2018optimal}, when the outcome for the first subsystem is $
    \Pi(\theta) = \begin{psmallmatrix}
    \sin^{2}(\theta) & \cos(\theta)\sin(\theta) \\
    \cos(\theta) \sin(\theta) & \cos^{2}(\theta)
    \end{psmallmatrix}
$,
the expected success probability given optimal choice of subsequent measurement is given by
\begin{align*}
\max_{\{\hat{\Pi}^{(2)}(d_{1})  \}} \Bigg( \text{P}_{\text{succ}} \Big( \hat{\Pi}^{(2)}(\theta)  \Big| \text{out}_{1} = \Pi(\theta)  \Big) \Bigg) & \leq \max_{\theta} \Big(\frac{1}{3} - \frac{1}{12}\cos(2\theta) - 0.288675 \cos(\theta) \sin(\theta) \\
    & + \frac{1}{2} \sqrt{\frac{5}{12} - \frac{1}{6} \cos(2 \theta) - 0.57735 \cos(\theta) \sin(\theta)}, \ 0.85 \Big) \\
    & = 0.933 \\
    &= \text{P}_{\text{succ}}\Big( \hat{\Pi}_{\text{AT}}, \big\{ \hat{\Pi}_{2}^{*}(d_{1}) \big\}  \Big).
\end{align*}
This proves that the best locally optimal strategy is the anti-trine with a success probability of $P_{\text{loc}}(\{\rho_{j}\}, \mathbf{q}) = 0.933$. Clearly, $\text{P}_{\text{loc}}(\{\rho_{j}\}, \mathbf{q}) < \text{P}_{\text{coll}}(\{\rho_{j}\}, \mathbf{q})$, from which it follows that a necessary condition for optimal locally adapative state discrimination of pure states is that $m=2$.  

\section{Proof of Theorem 2}
Any adaptive protocol consists of a series of measurements, $\{\Pi_{1}, \Pi_{2}(d_{1}), ... \Pi_{n}(\mathbf{d}_{[n-1]}) \}$, where all measurements after the first depend on previous measurement results. Then any individual measurement sequence can be written as a tensor product
\begin{align*}
    \Pi_{\mathbf{d}_{[n]}} &= \bigotimes_{k=1}^{n} \Pi_{k}(\mathbf{d}_{[k-1]}).
\end{align*}

Let $\mathcal{S}_{j}$ be the set of all measurement sequences $\mathbf{d}_{[n]}$ such that the post-measurement decoding is $\hat{\rho} = \rho_{j}$. Then, we can define $\Pi'_{j} = \sum_{\mathbf{d}_{[n]} \in \mathcal{S}_{j}} \Pi_{\mathbf{d}_{[n]}}$ as the measurement element which leads to decoding $\hat{\rho} = \rho_{j}$, and the difference between the two success probabilities can be bounded with

\begin{align*}
 \text{P}_{\text{succ}}\Big( \{\rho_{j}\} \Big) - \text{P}_{\text{succ}}\Big( \{\tilde{\rho}_{j}(\theta)\} \Big) &= \sum q_{j} \text{Tr}[\Pi'_{j} (\rho_{j} - \tilde{\rho}_{j}(\theta))] \\
 & \leq \max_{j} \Bigg(  \text{Tr}\big[\Pi'_{j} \big( \rho_{j} - \tilde{\rho}_{j}(\theta)\big) \big] \Bigg). 
\end{align*}

Then, by H{\"o}lder's inequality, we see that
\begin{align*}
    \left|\text{Tr}\Big[\Pi'_{j} \bigotimes_{k} \rho_{j}^{(k)}\Big] -  \text{Tr}\Big[\Pi'_{j} \bigotimes_{k} \tilde{\rho}_{j}^{(k)}(\theta)\Big] \right|
    & \leq \Big\| \Pi'_{j} \Big\|_{\infty} \Big\| \bigotimes_{k} \rho_{j}^{(k)} - \bigotimes_{k} \tilde{\rho}_{j}^{(k)} \Big\|_{1} \\
    & \leq \Big\| \bigotimes_{k} \rho_{j}^{(k)} - \bigotimes_{k} \tilde{\rho}_{j}^{(k)} \Big\|_{1}, 
\end{align*}
where the last inequality follows from noting that $\Pi_{j}' \leq \mathbb{I}$. Then we find that
\begin{align*}
    \Big\| \bigotimes_{k=1}^n \rho_{k} - \bigotimes_{k=1}^n \tilde{\rho}_{k} \Big\|_{1}
    & =  \Big\| \bigotimes_{k=1}^n \rho_{k} - \rho_{1} \otimes \bigotimes_{k=2}^n \tilde{\rho}_{k}  + \rho_{1} \otimes \bigotimes_{k=2}^n \tilde{\rho}_{k} -  \bigotimes_{k=1}^n \tilde{\rho}_{k} \Big\|_{1} \\
    & \leq \Big\| \bigotimes_{k=1}^n \rho_{k} - \rho_{1} \otimes \bigotimes_{k=2}^n \tilde{\rho}_{k} \Big\| + \Big\| \rho_{1} \otimes \bigotimes_{k=2}^n \tilde{\rho}_{k} -  \bigotimes_{k=1}^n \tilde{\rho}_{k} \Big\|_{1} \\
    & \leq \sum_{\ell = 0}^{n-1} \Big\| \bigotimes_{k=1}^{\ell+1} \rho_{k} \otimes \bigotimes_{k=\ell+2}^{n} \tilde{\rho}_{k} - \bigotimes_{k=1}^{\ell} \rho_{k} \otimes \bigotimes_{k=\ell+1}^{n} \tilde{\rho}_{k} \Big\|_{1} \\
    & = \sum_{\ell = 0}^{n-1} \Big\| \bigotimes_{k=1}^{\ell} \rho_{k} \otimes \Big( \rho_{\ell+1} - \tilde{\rho}_{\ell+1} \Big) \otimes \bigotimes_{k=\ell+2}^{n} \tilde{\rho}_{k} \Big\|_{1}.
\end{align*}
Let us consider the $\ell^{\text{th}}$ term. We denote the eigenvalues of $\rho_{k}$ as $\{\lambda_{1}^{(k)}, \lambda_{2}^{(k)} \}$ and likewise the eigenvalues of $\tilde{\rho}_{k}$ as $\{\tilde{\lambda}_{1}^{(k)}, \tilde{\lambda}_{2}^{(k)}\}$. Finally, we denote the eigenvalues of $\rho_{\ell+1} - \tilde{\rho}_{\ell+1}(\theta)$ as $\{\sigma_{1}(\theta), \sigma_{2}(\theta)\}$. Then, we have 
\begin{align*}
    \Big\| \bigotimes_{k=1}^{\ell} \rho_{k} \otimes \Big( \rho_{\ell+1} - \tilde{\rho}_{\ell+1} \Big) \otimes  \bigotimes_{k=\ell+2}^{n} \tilde{\rho}_{k} \Big\|_{1} &= \prod_{k=1}^{\ell} \Big( |\lambda_{1}^{(k)}| + |\lambda_{2}^{(k)}| \Big) \times \Big(|\sigma_{1}(\theta)|+|\sigma_{2}(\theta)|  \Big) \\
    & \hspace{4em} \times  \prod_{k=\ell+2}^{n} \Big( |\tilde{\lambda}_{1}^{(k)}| + |\tilde{\lambda}_{2}^{(k)}| \Big) \\
    &= |\sigma_{1}(\theta)|+|\sigma_{2}(\theta)|
\end{align*}
since $|\lambda_{1}^{(k)}| + |\lambda_{2}^{(k)}|=1$ for all $k$. Now, we bound the eigenvalues of $\rho_{\ell} - \tilde{\rho}_{\ell}$ with
\begin{align*}
    \Big\| \rho_{\ell} -\tilde{\rho}_{\ell} \Big\|_{1}
    & = \Big\| \rho_{\ell} -  U(\theta) \rho_{\ell} U(\theta)^\dagger \Big\|_{1} \\
    & = \Big\| \Big[\rho_{\ell},  U(\theta)  \Big] U(\theta)^\dagger \Big\|_{1} \\
    & = \Big\| \Big[  \rho_{\ell} - \mathbb{I}, U(\theta) - \mathbb{I} \Big] \Big\|_{1}, 
\end{align*}
where we have used that identity commutes with all operators and the unitary invariance of the trace norm. Since $0 \leq \rho_{\ell} \leq \mathbb{I}$ all $\ell$ we have that
\[
\Big\| \rho_{\ell} - \tilde{\rho}_{\ell} \Big\|_{1} \leq \Big\| U(\theta) - \mathbb{I} \Big\|_{1}.
\]
Since $U(\theta)$ is a rotation, we can easily calculate its eigenvalues to $e^{\pm i \theta}$. Hence, we conclude that 
\begin{align*}
    \Big\| \rho_{\ell} - \tilde{\rho}_{\ell} \Big\|_1 & \leq \Big|e^{i \theta} - 1\Big| + \Big| e^{-i \theta} - 1\Big| \\
    & = 4 \sin(|\theta|/2).
\end{align*}
From this, we have 
\begin{align*}
     \Big\| \bigotimes_{k=1}^n \rho_{k} - \bigotimes_{k=1}^n \tilde{\rho}_{k} \Big\|_{1} \leq \sum_{\ell = 0}^{n-1} 4 \sin(|\theta|/2) = 4n \sin(|\theta|/2)
\end{align*}
and the statement follows.

\end{document}